\def\year{2021}\relax
\documentclass[letterpaper]{article} 
\usepackage{aaai21}  
\usepackage{times}  
\usepackage{helvet} 
\usepackage{courier}  
\usepackage[hyphens]{url}  
\usepackage{graphicx} 
\usepackage{natbib}  
\usepackage{caption} 
\usepackage{subcaption}
\usepackage{makecell}
\usepackage{tabularx}
\usepackage{amsmath}
\usepackage{multirow}
\usepackage{paralist}
\newcommand{\norm}[1]{\left\lVert#1\right\rVert}
\newcommand{\cmttfont}[1]{\fontfamily{cmtt}\selectfont}
\DeclareTextFontCommand{\textcmttfont}{\cmttfont}
\title{Divided We Rule: Influencer Polarization on Twitter During Political Crises in India}
\author{
    Saloni Dash,\textsuperscript{\rm 1}
    Dibyendu Mishra,\textsuperscript{\rm 1}
    Gazal Shekhawat,\textsuperscript{\rm 1}
    Joyojeet Pal\textsuperscript{\rm 1}
    \\
}
\affiliations{

    \textsuperscript{\rm 1} Microsoft Research
    \\ Bengaluru, India \\
    
}

\begin{document}

\maketitle

\begin{abstract}

Influencers are key to the nature and networks of information propagation on social media. Influencers are particularly important in political discourse through their engagement with issues, and may derive their legitimacy either solely or in large part through online operation, or have an offline sphere of expertise such as entertainers, journalists etc. To quantify influencers' political engagement and polarity, we use Google's Universal Sentence Encoder (USE) to encode the tweets of 6k influencers and 26k Indian politicians during political crises in India. We then obtain aggregate vector representations of the influencers based on their tweet embeddings, which alongside retweet graphs help compute their stance and polarity with respect to these political issues. We find that influencers engage with the topics in a partisan manner, with polarized influencers being rewarded with increased retweeting and following. Moreover, we observe that specific groups of influencers are consistently polarized across all events. We conclude by discussing how our study provides insights into the political schisms of present-day India, but also offers a means to study the role of influencers in exacerbating political polarization in other contexts.

\end{abstract}

\section{Introduction}
As social media raises concerns of political polarization worldwide, questions arise on the role of public figures who wield the ability to influence the political discourse. In parts of the west, there are inherent tensions between what are largely seen as liberal mainstream celebrities
\cite{demaine2009navigating} and an emerging group of hyper-partisan micro-celebrities \cite{lewis2020news}. In India, the role of influencers is complex and often has regional ramifications. Mainstream celebrities tend to support the incumbent central leadership, which in general is organized online \cite{lalani2019appeal, mishra2021rihanna}, while public figures speaking against the government have faced online harassment, trolling, and attacks to their livelihoods \cite{singh2020politics, menon2020hindu}.
\newline \indent
In this paper, we study the role of influencers in amplifying political polarization in India, by quantifying the partisan engagement of Twitter handles of Indian entertainers, sportspersons, journalists and other public figures alongside handles of politicians from two major national parties in India -- the national incumbent BJP (Bharatiya Janata Party) and the opposition party INC (Indian National Congress). We do this by examining influencer engagement on four issues that have dominated the news and seen divisive political activism during June 2019 - March 2021. In the order of their onset, they are as follows --
\newline
\noindent \textbf{Abrogation of Article 370:} The central government’s unilateral abrogation of Article 370 of the Indian constitution, which had secured autonomy for India’s only Muslim majority state, Jammu \& Kashmir was followed by the controversial communications ban, detainment of politicians, a curfew, and increased armed forces on the ground.
\newline
\noindent \textbf{Citizenship Amendment Act (CAA):} The Act, which disqualified Muslims from qualifying for refugee status when entering from India’s neighboring countries, was enacted alongside the government's plans to conduct a nationwide exercise to identify ``illegal foreigners" living in India by requiring historical documentation of residency. 
This stoked fears of losing citizenship among India’s Muslims and led to widespread protests all around the country.
\newline 
\noindent \textbf{COVID-19:} 
The ``first wave'' period was marked by the government announcing a nationwide lockdown starting March 25, 2020, triggering an exodus of low-income migrant workers from cities to hinterlands, often on foot, with transport lines closed. The migrant crisis and economic impacts and health concerns and practices were part of the contentious issues in these conversations.
\newline 
\noindent \textbf{Farmers' Protests:} The Indian Farm Bills which were passed in parliament in September 2020, targeted pricing and  subsidies and allowed the entry of corporations into certain domains of crop trading. This led to protests by farmers, particularly from the states of North India, where the new configurations of crops and purchase prices raised concerns of threats to their livelihoods.
\newline \indent
The four events we study here bear elements of communication and outreach both for and against the government's position. While three events are related to responses on specific government initiatives (Article 370, CAA, Farm Bills), the COVID-19 related discourse is an ongoing issue with longer term sustained social media communication. 
Figure 1 shows a weekly timeline (June 2019 - March 2021) of influencers retweeting politicians, with peaks corresponding to the four events being studied. We see that influencer retweets on both sides tend to spike around the same events with the exception of a few events, such as \textit{Howdy Trump} when the ruling party mobilized more influencers, while influencers on the opposition side mostly ignored it. 

\begin{figure}[!htbp]
  \centering
 \includegraphics[width=0.99\columnwidth]{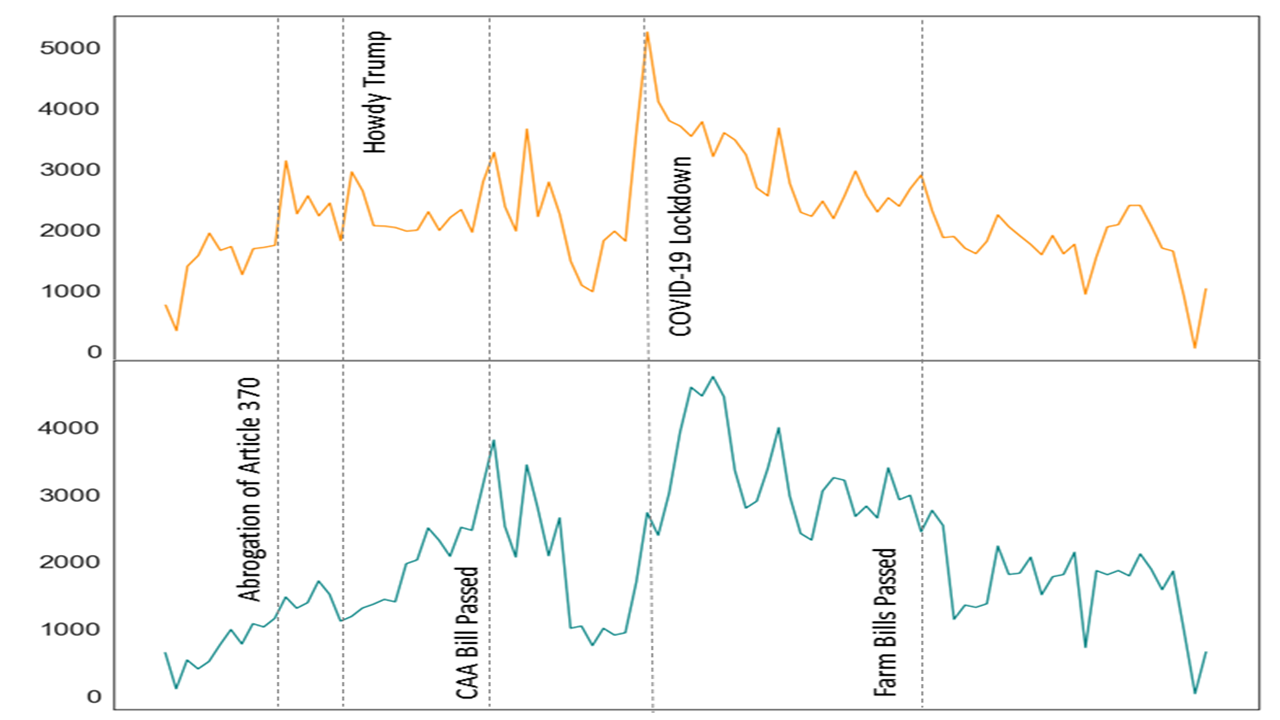}
  \caption{\textbf{Influencers' Timeline of Retweeting BJP (Top) and INC (Bottom) Politicians.} 
}
\end{figure}

\begin{table}[!htbp]
\centering
\small
\resizebox{\columnwidth}{!}{
\begin{tabular}{|c|c|c|c|c|c|}
\hline
\multirow{2}{*}{\textbf{Party}} & \multicolumn{4}{c|}{\textbf{Retweet Range}} & \multirow{2}{*}{\textbf{Total}} \\
 & \multicolumn{1}{l|}{\textbf{{[}1,50{]}}} & \multicolumn{1}{l|}{\textbf{(50,100{]}}} & \textbf{(100,150{]}} & \textbf{(150,\textit{max}]} &  \\ \hline
BJP & 3265 & 214 & 99 & 272 (\textit{max}-10053) & 3850 \\ \hline
INC & 3144 & 185 & 62 & 151 (\textit{max}-54301) & 3542 \\ \hline
\end{tabular}}
\caption{\textbf{Distribution of Influencers Retweeting BJP and INC Politicians.} 
}
\label{influencers_rt_table}
\end{table}

In Table \ref{influencers_rt_table} we see that most individual influencers have generally engaged with politicians in small measure (1-50 times). However, we see that a significant number of influencers have engaged with politicians' content more than 150 times, with a single influencer retweeting INC politicians 54k times. This concentration of influencers on the high end suggests the existence of a loyal set of accounts that actively promote a party's political content.
\newline \indent 
Our paper investigates political engagement of Indian influencers through a computational lens by addressing following questions -- 
a) \textit{Do influencers engage with politically polarizing issues in a partisan manner?}
b) \textit{Are politically polarized influencers rewarded for their partisan engagement?}
c) \textit{Are certain groups of influencers (journalists, entertainers etc.) consistently polarized towards a political party across all the issues?}

We propose a novel workflow where we collect tweets of 26k politicians and 6k influencers and filter them to obtain event-specific tweets. We then use the tweets to compute the stance of the influencers for each of the events, which is further used to quantify their partisan engagement. Finally, we use the computed partisan scores to identify and characterize polarized influencers. Our contributions include --
\begin{compactitem}
\item We collect over 43M English tweets of politicians \& influencers and use a a Word2Vec \cite{mikolov2013efficient} based technique to classify them.
\item Using the tweets, we compute the stance of influencers for each of the events to find if influencers engage on these topics in a partisan manner. We thereafter identify dominant polarizing narratives that they engage in.
\item We then introduce polarity metrics to quantify the partisan engagement by using the content of the tweets as well as the induced retweet network.
\item The polarity scores are further used to characterise partisan influencers, who, we find, are significantly retweeted and followed in comparison to other users. Moreover, certain categories of influencers, including the emerging category of platform celebrities, are consistently polarized across all events.
\item We conclude by discussing the implications of our findings, including the significance of issue-based polarization in other multi-ethnic democracies like India.
\end{compactitem}

\section{Related Work}
Two interlinked strands of literature on communication networks on Twitter are of importance to our study: the question of political polarisation and echo chambers on social media, as well as literature on influencers and their role in political communications online.
\subsection{Political Polarization} 
In  their  formative  study  of  political  polarization,  \citet{conover2011political} explain homophily on social media through partisan user behavior in the United States. The authors find the retweet network of their sample to be highly polarized, with limited interaction between the two camps. Since the initial studies on Twitter relating to echo chambers, the debate has persisted on whether they are overstated \cite{mukhudwana2020zuma}, or merely reflective of offline divisions \cite{siapera2015gazaunderattack}.

As \citet{recuero2020hyperpartisanship}’s work on Brazil has shown, the causation of partisanship online and offline need not be mutually exclusive. In the backdrop of the rise of populist movements across USA, Hungary, Philippines and India, the authors capture the asymmetrical nature of polarisation, and found that the pro-Bolsonaro camp in Brazil fostered a hyper-partisan environment online. They found that in such scenarios of disproportionate polarisation, biased information spreads with ease, as partisan accounts and media houses that defy traditional notions of detachment from the contested subject gain prominence \cite{recuero2020hyperpartisanship,faris2017partisanship}. Similarly, \citet{neyazi2016campaigns} find Twitter networks in India to be highly polarizing,  especially in BJP’s political campaigns.

An inordinate focus on American case studies and political events \cite{barbera2015understanding} risks affirming the impression of a Twitter base that is uniform and continual in its divisions. Some works have reflected on gaps in single-issue based studies of polarisation. For instance, cross-topic research on controversies in Russia, the USA and Germany found that users do not necessarily diverge on the camps of ``left" and ``right" that have become synonymous to popular studies of polarisation that emerge from the USA. Instead, issue-based publics reflect multiple combinations of stances across the right-left dichotomy\cite{bodrunova2019beyond}. 

In the Indian case, divisions on the basis of caste, religion or linguistics have had a historical continuity, however, which intersections translate to online fissure or coalescence remain unclear. Modi’s own brand of populist politics remains close to Hindu nationalism, yet, in considering four different events, we extend our analysis to controversies both central and peripheral to the realm of Delhi-based politics.  In doing so, we intend to further our understanding of the prominence of issue based rhetoric within the region, across events. While the axis of polarisation is still between the BJP and INC, we undertake a ground truth analysis to determine whether polarisation towards one party translates to vocal support of its policies, or if it is simply, dissent against the other party’s role in disparate events.



\subsection{Influencer Engagement on Social Media}
Apart from the question of divergences in online user bases, the question of how such a polarisation is achieved or influenced is fundamental to understanding political polarization on Twitter. A body of work suggests that influencers exacerbate polarization on social media by endorsing and spreading partisan content. In an experiment with Democrats and Republicans, \citet{becker2019wisdom} demonstrated that polarization does not spread in egalitarian networks where all individuals have equal influence throughout the network. They argue that partisan bias is amplified on social media platforms like Twitter because these networks are organised around few key influencers \cite{centola_2020} ,
and the ability to obtain central positions could allow extremists to generate and sustain belief polarization in social media networks \cite{becker2019wisdom}.
Supporting this is \citet{garibay2019polarization}, who find that polarization in social media helps influencers gain more followers. 

There also exists a robust tradition of literature that focuses on the impact and interactions of politicians with the media and journalists. In  India,  studies  of  Modi’s  political  campaigning  have particularly noted the use of celebrities on Twitter for image building \cite{pal2015banalities}. By interacting with key influencers, Modi  was  able  to  successfully  attach  his  brand  with  that of  leading  entertainers  and  sportspersons. The other category of accounts that have come to the limelight for attempting to influence political discourse on Twitter are those of ``trolls" and ``bots" \cite{zannettou2019let,stewart2018examining}, which may be automated accounts that tow a partisan line and tend to spread unreliable information. However, it may be difficult to determine whether an account is human or programmed, or if trolls are individuals compensated for their disruption of interactions.  

Recent work points to a decline in the influence of traditional media houses and sources of celebrity \cite{bodrunova2016influencers, grave2017exploring}, making the study and theorizing around online influencers in the Indian context important. We devise the category of ``platform celebrity" to describe accounts who owe their popularity to their online presence, rather than their offline occupation. While this includes some YouTubers or Twitter users whose commentary or wit goes viral frequently, the influencers coded under this category rarely share  insights about their offline lives or experiences. Instead, they engage in partisan commentary on the latest controversy. Further, we diverge from numerical understandings of describing influence \cite{jiang2021building}, to create a context specific typography of Twitter influencers based on their occupational backgrounds. In doing so, we hope to disaggregate the notion of influence as it relates to online political discourse, to pave the way for understanding localised forms of polarisation in different regions of the world. 

Existing studies closest to our work include \citet{jiang2021social}, who explore political polarization in USA during COVID-19. This proposes Retweet-BERT, a model that uses profile descriptions and retweet networks to estimate the polarity of a user. However, the study considers general tweets on COVID-19, while we focus solely on messaging by individuals who are known to wield public influence online or offline. Besides, we also use the tweets of the users to compute their stance on the issue, which offers more information on ideological leaning than profile description. Moreover, we use an embedding-based unsupervised algorithm to detect their stance, whereas \citet{jiang2021social} use a weakly-supervised method to identify and quantify the stance, which is often not feasible due to the lack of labelled data. 

\section{Dataset}
We collect English tweets from 26,435 BJP and INC politicians and 6,626 Indian influencers. We manually classify the influencers into 12 categories, based on their primary area of influence. The details are provided below.
\subsection{Politicians}
We use a publicly available dataset \cite{panda2020nivaduck} of Twitter accounts of 36k+ Indian politicians which include elected representatives from various parties at the state and national level, as well as volunteers like general secretaries, spokespersons etc. Briefly, the dataset is curated by using a scalable Machine Learning classification pipeline called NivaDuck, and further validated manually by human coders. Moreover, the politicians are also manually annotated by the state and party they belong to. We use these party labels to filter 14,094 BJP and 12,342 INC politician accounts. We then collect tweets
\footnote{Using Tweepy - \url{https://www.tweepy.org/}} from these 26,435 accounts between June 2019 - March 2021.

\subsection{Influencers}
To build the set of influencers we use the Twitter API to collect the friends of the filtered politicians, i.e. accounts that the politicians follow, with the assumption that politicians follow other politicians as well as public figures such as journalists and celebrities. This results in over 100k+ accounts from which we remove accounts that are not followed by at least three users from the initial set of politicians. This further reduces the annotation set to include accounts that are influential enough to be followed by a certain threshold of politicians. The threshold can be increased to make the filtering more conservative. Additionally, we remove all politicians from the NivaDuck set \cite{panda2020nivaduck} and non-Indian global figures, such that we are left with 10k Twitter accounts of potential influencers that are highly followed by Indian politicians. 
\newline
\indent It is worth noting that since our process is built ground up from the friends of politicians, it implies that our definition of influencer is biased towards those accounts that are of interest to politicians - i.e. for instance, if our seed set had started with sportspersons, it is likely we would have ended up with several highly followed sports management companies or commentators in our set, or likewise for other domains. In essence, a journalist with 3000 followers, of which 200 followers are politicians would be included in our set, as opposed to say a PR agent with 300k followers, but with few politicians following them.
\newline \indent
However, the team's contextual knowledge of India is used to ensure that known public figures are included in the sample as far as possible. Thus, we are confident that the vast majority of ``A List" public figures such as film stars, sportspersons, businesspersons etc. are included in our sample, but as Table 2 shows, we capture a relatively high number of journalists and media houses, since their work is of obvious importance to politicians. We further manually categorize every influencer account into one category, whichever they are primarily known by -- thus a major sportsperson with business interests is nonetheless categorized under sports.

\begin{table}[!htb]
\centering
\begin{tabular}{|l|l|}
\hline
\textbf{Category}                    & \textbf{Count}            \\ \hline
Academia                             & 172                       \\ \hline
Activist                             & 81                        \\ \hline
Business                             & 208                       \\ \hline
Entertainment                        & 1251                      \\ \hline
Fan Account                          & 36                        \\ \hline
Journalist                           & 3551                      \\ \hline
Law \& Policy                        & 141                       \\ \hline
Media House                          & 550                       \\ \hline
Platform Celebrity                   & 126                       \\ \hline
Social Work                        & 63                        \\ \hline
Sports                               & 348                       \\ \hline
Writer                               & 99                        \\ \hline
\multicolumn{1}{l}{\rule{0pt}{2.6ex}\textit{Total}} & \multicolumn{1}{l}{\textit{6626}} \\ 
\end{tabular}
\caption{\textbf{Influencers by Category}}
\label{inf_cat}
\end{table}

The final set of influencers are divided into a 12 categories. \textit{Academia} refers to accounts that largely work with higher education institutes and conduct research. The \textit{Activist} category includes grassroots organizers, leaders or prominent members of nonprofits, affiliates of religious or quasi-religious organizations who present themselves as activists, or any who self-identify with the label. The \textit{Business} category includes accounts of commercial establishments, their founders, and key employees. The \textit{Entertainment} segment refers to people engaged in films, television, music, and fashion. \textit{Fan Accounts} have to do with profiles that distinctly avow to support public figures or organizations, and includes unofficial accounts dedicated to actors, politicians and political parties. \textit{Journalist} refers to individuals who work in the reporting and production of news, features and commentary. The \textit{Law \& Policy} category is for lawyers, judges, and public policy practitioners. 
\textit{Media Houses} are platforms, both digital and conventional that produce news, entertainment etc and thus this category also includes newspapers, websites, and TV channels. Like \textit{Fan Accounts}, the \textit{Platform Celebrity} category captures a digital phenomenon of accounts whose popularity stems from their online presence and content creation, rather than the offline work of the individuals involved. The \textit{Social Work} category refers to the non-profit sector and includes accounts of its employees. \textit{Sports} captures both athletes, as well as the accounts of teams, tournaments, and governing bodies. \textit{Writer} is a category for published authors of fiction and non-fiction, but does not include columnists employed by news platforms.

\subsection{Pre-Processing}
Tweets often contain hyperlinks, emojis, hashtags etc. which  can impede computational analysis. All tweets are thus pre-processed in the following manner --
\begin{compactenum}
\item Removing all hyperlinks.
\item Removing emojis using the python emoji package \cite{kimemoji}.
\item Removing retweet (RT @) and user (@) mentions.
\item Removing punctuations \& non-alphanumeric characters.
\item Case folding letters to lower case.
\end{compactenum}
\section{Tweet Classification}
In order to capture event-specific tweets, we use a Word2Vec \cite{mikolov2013efficient} bag of words based technique to classify the collected tweets \cite{vijayaraghavan2016automatic}. We first curate a set of high precision keywords that are indicative of the event. We define high precision keywords as words that are most likely used only in the context of the event, i.e. for Farmers' Protests we use \textcmttfont{\#farmersprotests} instead of just \textcmttfont{farmer} which can be used in other contexts, such as increasing farmer suicide rates or other agricultural initiatives. 

We then filter tweets that contain at least one of the keywords, and train a Word2Vec model on the filtered tweets. After obtaining a vector representation of all the words, we expand each of the keywords from the seed set, by computing the cosine similarity of pairs of words for all words in the vocabulary. We iteratively add words to the seed set according to the cosine similarity based criteria defined in \citet{vijayaraghavan2016automatic}. Some of the keywords and hashtags are showed in Table \ref{topic_keywords}.

\begin{table}[!htb]
\centering
\small
\begin{tabular}{p{0.3\columnwidth}|p{0.65\columnwidth}}

        & \textbf{Keywords \& Hashtags} \\
        \hline
\rule{0pt}{2.6ex}\textbf{Article 370} &  
\textcmttfont{\#jammuandkashmir}, \textcmttfont{\#article370}, 35a, abrogation, kashmiri \\

\textbf{CAA/NRC} &  \textcmttfont{\#caanrc},\textcmttfont{\#caanrcnpr},\textcmttfont{\#anticaa}, \textcmttfont{\#caanrcprotest},
\textcmttfont{\#citizenshipamendmentact}   \\
\textbf{COVID-19} & covid2019, coronavirus,\textcmttfont{\#covidpandemic}, \textcmttfont{\#coronapandemic}, \textcmttfont{\#coronacrisis}  \\
\textbf{Farmers' Protest} & \textcmttfont{\#farmersprotests}, \textcmttfont{\#farmersagitation},\textcmttfont{\#delhichalo}, \textcmttfont{\#farmersdelhiprotest}
\end{tabular}
\caption{\textbf{Keywords \& Hashtags by Event.}}
\label{topic_keywords}
\end{table}
To obtain a rough estimate of the performance of our classification pipeline, we randomly select a representative sample (i.e. we calculate sample size using 95\% confidence level and 3\% confidence interval) and have two annotators manually annotate whether each tweet is related to the particular event or not. We report precision values of 75.5\%, 99.6\%, 99.7\% and 93.7\% for Article 370, CAA/NRC, COVID-19 and Farmers' Protests respectively. Overall, we obtain a combined precision of 92.1\% with an average inter-annotator agreement of 98.7\%.

\begin{table}[!htb]
\centering
\small
\begin{tabular}{|c|c|c|}
\hline
\textbf{Topic} & \textbf{Users}  & \textbf{Tweets} \\
\hline
Article 370 &   14,050        & 346,197 \\
\hline
CAA/NRC & 13,744 & 527,702 \\
\hline
COVID-19 & 20,871 & 3,069,851 \\
\hline
Farmers' Protest & 12,224 & 269,742 \\ \hline

\end{tabular}
\caption{\textbf{Users and Tweets by Event}}
\label{topic_stats}
\end{table}
The number of tweets for each event, identified by the classification pipeline, including the number of users who have tweeted about that particular event are in Table \ref{topic_stats}. 

\section{Partisan Engagement of Influencers}
We use the event-specific tweets of influencers to detect their position on each of the polarizing issues. We use the method proposed by \cite{rashed2020embeddings}, which we briefly describe below. We find that overall, the stance of the influencers always aligns with that of one of the political parties, suggesting that influencers engage on these polarizing issues in a partisan manner. We also analyse the prominent narratives from both sides of the issue to further contextualise the partisan engagement. 
Additionally, we quantify this polarized interaction using metrics based on the retweet network \cite{garimella2018quantifying} and content of the tweets \cite{waller2020community}.

\subsection{Stance Detection Using Tweet Embeddings}
Given a set of accounts and their tweets for an event, we obtain user embeddings based on aggregate tweet embeddings \cite{rashed2020embeddings}. We use Google's Universal Sentence Encoder (USE) \cite{cer2018universal} to obtain high dimensional vector representations of the tweets by a user for a given event, and average out all the tweet embeddings to represent a single user. The user embeddings are projected to a two dimensional space using UMAP \cite{mcinnes2020umap} and clustered using hierarchical density based clustering (HDB-SCAN) \cite{mcinnes2017accelerated}.

\begin{figure}[!htb]
\centering
\begin{subfigure}{.5\columnwidth}
  \centering
  \includegraphics[width=0.99\linewidth]{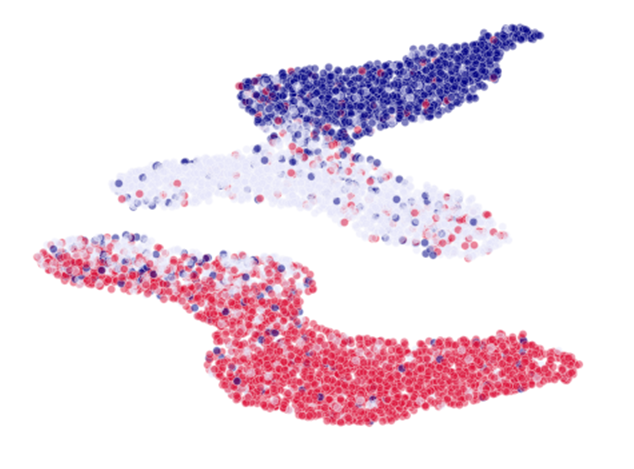}
  \caption{Article 370}
  \label{art370_cluster}
\end{subfigure}%
\begin{subfigure}{.5\columnwidth}
  \centering
 \includegraphics[width=0.99\linewidth]{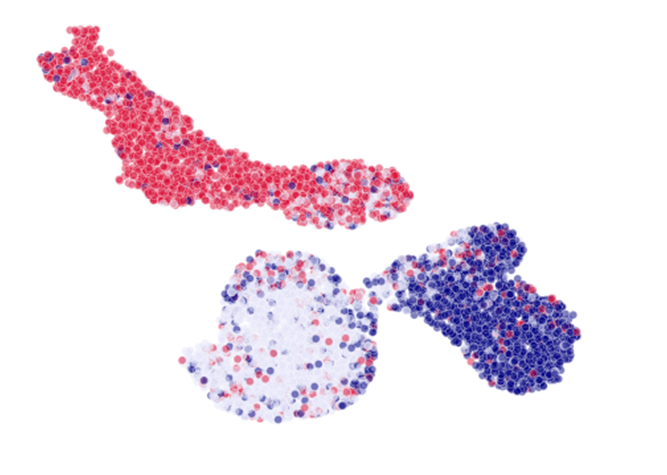}
  \caption{CAA/NRC}
  \label{caanrc_cluster}
\end{subfigure}
\begin{subfigure}{.5\columnwidth}
  \centering
 \includegraphics[width=0.99\linewidth]{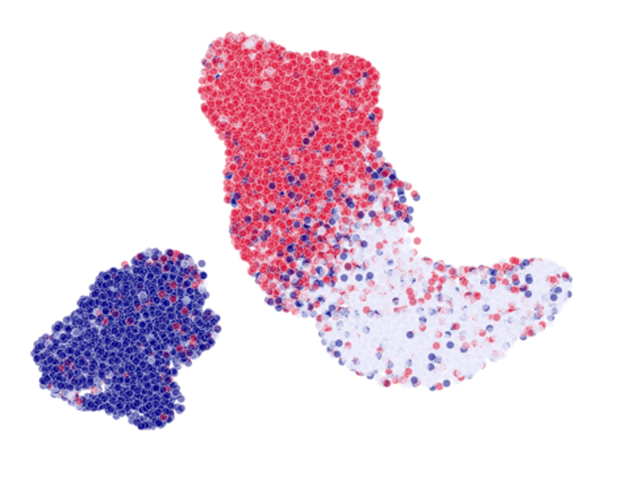}
  \caption{COVID-19}
  \label{covid_cluster}
\end{subfigure}%
\begin{subfigure}{.5\columnwidth}
  \centering
 \includegraphics[width=0.99\linewidth]{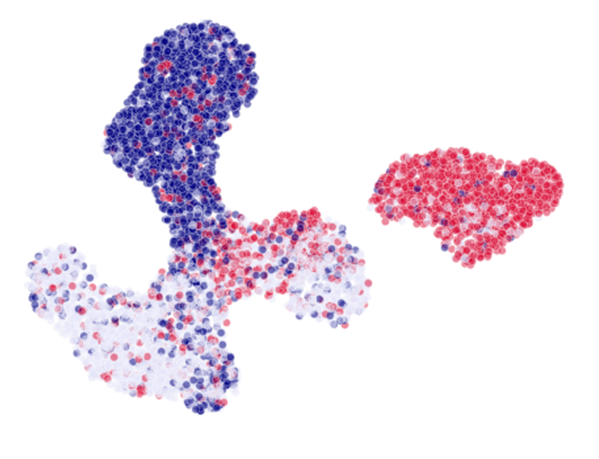}
  \caption{Farmers' Protests}
  \label{farmers_cluster}
\end{subfigure}
\begin{subfigure}{.65\columnwidth}
  \centering
 \includegraphics[width=0.9\linewidth]{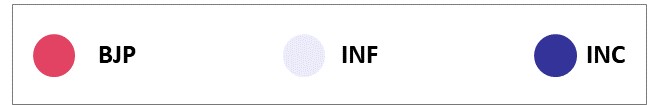}
  \caption{}
  \label{legend}
\end{subfigure}
\caption{\textbf{HDB-SCAN Clusters by Event}. Influencers are clustered according to their stance on the issue.}
\label{topic_clusters}
\end{figure}

As shown in Figure \ref{topic_clusters},
the users are clustered according to their stance. We see two clusters across all the events, one for each of the stances (pro-INC and pro-BJP). We see that INC and BJP politicians lie in separate clusters across all the events, which reaffirms the polarizing nature of the events. Moreover, we observe that the influencers (INF) also lie in either of the two clusters, confirming our hypothesis that the influencers engage with these issues in a partisan manner. We observe that for all of the events except COVID-19, majority of the influencers are in the pro-INC cluster, whereas for COVID-19 majority of them are in the pro-BJP cluster. 
\newline \indent 
To understand this pattern further, we manually studied the individual tweets. The overall clustering of influencers towards the INC is explained by the engagement of journalistic content by the opposition party, and the reliance of the ruling party on relatively more influential, but less numerous platform celebrities, or mainstream celebrities. Thus while the overall number of influencers seem to lie in the INC more, this finding needs to be seen alongside the retweet polarity scores (Figure \ref{rt_polarity_cat}) which help understand the size and scope of influencer engagement. Moreover, we observe that the relatively larger clustering on the INC side is a result its politicians' tendency to engage with influencers on account of their criticism of the BJP -- this does not mean those individuals are politically aligned with the INC, simply that they are critical of the ruling party. The BJP's engagements and the resultant clusters, on the other hand, are closest to accounts who advocate for the party. 
\newline \indent
These patterns speak to the larger state of politics in India,  where political opposition on a national level is weak, and civil society members and journalists, rather than the INC, have emerged as popular critics of the government. Our section on categorical analysis of the influencers further explains the distinctiveness of the COVID-19 clustering.

\subsection{Polarized Narratives}
\begin{figure}[!htb]
\centering
\begin{subfigure}{.75\columnwidth}
  \centering
  \includegraphics[width=0.99\linewidth]{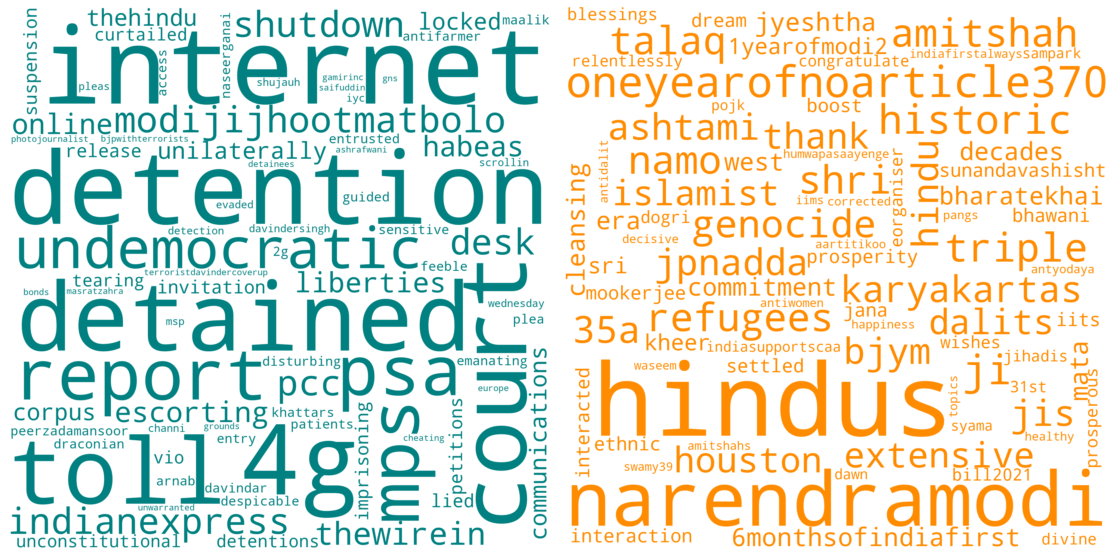}
  \caption{Article 370}
\end{subfigure}
\begin{subfigure}{.75\columnwidth}
  \centering
 \includegraphics[width=0.99\linewidth]{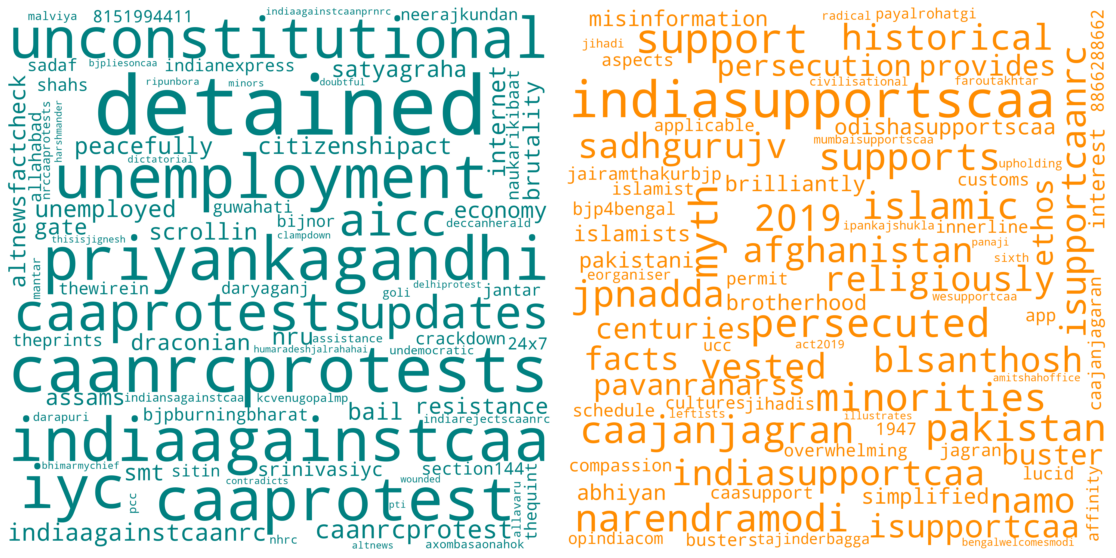}
  \caption{CAA/NRC}
\end{subfigure}
\begin{subfigure}{.75\columnwidth}
  \centering
  \includegraphics[width=0.99\linewidth]{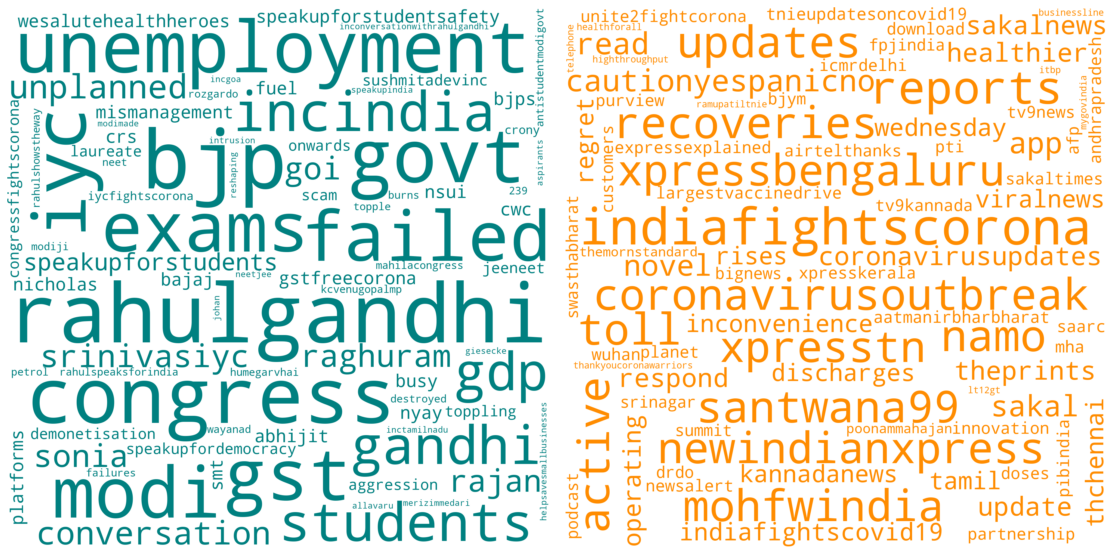}
  \caption{COVID-19}
\end{subfigure}
\begin{subfigure}{.75\columnwidth}
  \centering
 \includegraphics[width=0.99\linewidth]{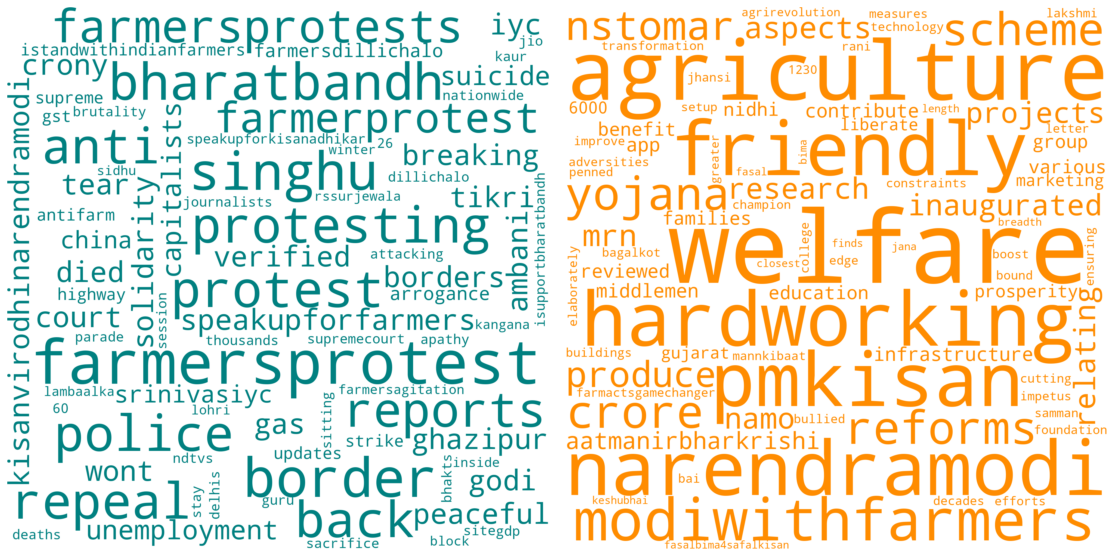}
  \caption{Farmers' Protests}
\end{subfigure}
\caption{\textbf{WordClouds of BJP (Right, in Orange) and INC (Left, in Green) Clusters by Event}}
\label{wordcloud}
\end{figure}

We further study the dominant narratives that influencers from each of the polarized clusters propagate, through the wordclouds in Figure \ref{wordcloud}. The wordclouds are generated using prominence scores \cite{rashed2020embeddings} derived from tweets of influencers from each cluster. Briefly, prominence scores calculate how often terms occur in the set of tweets from users of one cluster as compared to the other cluster. Therefore, the terms displayed in the wordclouds are those that are prominently used by influencers from each of the sides and thus help underscore the different narratives. 

On Article 370 the pro-INC side criticize the actions of the government in shutting down internet in Jammu \& Kashmir (J\&K) and the detention of several J \& K officials based on the Public Safety Act (PSA), while the pro-BJP side celebrate the abrogation, viewing it as a major step in resettling the Kashmiri pandits, who had fled the state due to terror campaigns inflicted on them by insurgents looking to establish Islamic rule.
The CAA/NRC issue saw the pro-INC side labelling the act as ``unconstitutional" while the pro-BJP tweets celebrate it as ``historic", having citizen support and hailing it as a humanitarian act addressing refugee-related issues. In essence, we see here shades of populism, where one side attempts to speak of the issue in legalistic terms, while another focuses on the partisan appeal of the act. 

On COVID-19, a less divisive issue (at the time), INC sympathizing influencers engage on ``unemployment" and ``failure", while on the side of the ruling party, the engagements are more utilitarian, focusing on the ``fight" against the Coronavirus.  
With the Farm Bills, we see a different form of populist engagement, in which the INC side uses terms such as ``draconian", and ``anti-farmer" while the BJP side highlights farmers as heroic using brand terms like ``aatmanirbhar" (self-reliant), ``hardworking" etc.  Unlike with CAA/NRC and 370, where there is a clear nativist appeal in creating an ``other" in those that oppose the act, the discourse with the farm bill is a bit different, since the ruling party would rather not demonize farmers, and must toe a careful line in countering opposition online.

\subsection{Quantifying Polarization}
We use the polarized clusters to compute the degree of polarization of influencers. Two aspects of polarization are considered -- retweet polarity which captures which ``side" of the issue is more likely to endorse the influencer, and content polarity which signifies the extent to which the influencers' stance resembles that of either party on a particular issue.  

Let $\mathbf{V}$ be the set of all users who have tweeted about a particular event. The partition for $\mathbf{V}$, $\mathbf{X}$ and $\mathbf{Y}$ is provided by the clusters from HDB-SCAN such that $\mathbf{X}$ is the set of users from the pro-INC cluster and $\mathbf{Y}$ is the set of users from the pro-BJP cluster.

\subsubsection{Retweet Polarity}
We construct an undirected retweet graph $\mathbf{G}$ with users in $\mathbf{V}$ as vertices. An edge $e_{ij}$ exists between users $v_i$ and $v_j \in \mathbf{V}$, if $v_i$ and $v_j$ have at least two retweets between them, i.e. if either $v_i$ has retweeted $v_j$ at least twice, or $v_i$ has retweeted $v_j$ at least once and vice versa or $v_j$ has retweeted $v_i$ at least twice, $\forall i,j = 1...|\mathbf{V}|$. We choose two retweets as our threshold based on the experiments by \citet{garimella2018quantifying} to avoid unreliable results. We then use the Random Walk Controversy (RWC) score \cite{garimella2018quantifying} to define retweet polarity.

\begin{figure}[!htb]
\centering
\begin{subfigure}{.5\columnwidth}
  \centering
  \includegraphics[width=0.99\linewidth]{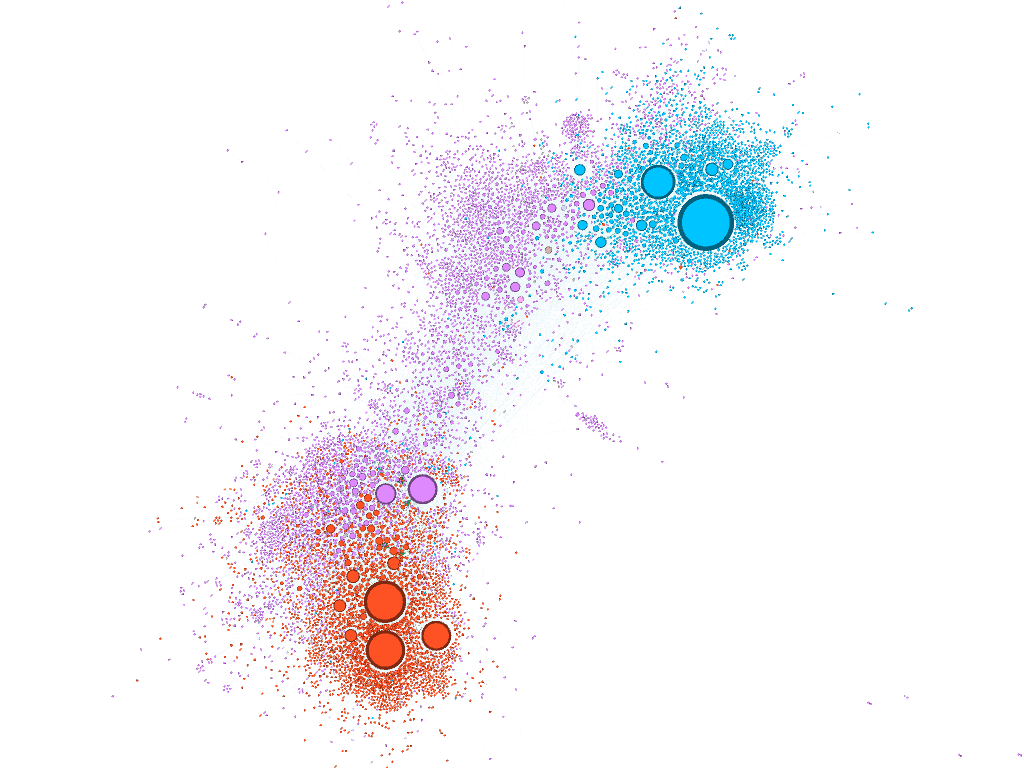}
  \caption{Article 370}
  \label{art370_rt_graph}
\end{subfigure}%
\begin{subfigure}{.5\columnwidth}
  \centering
 \includegraphics[width=0.99\linewidth]{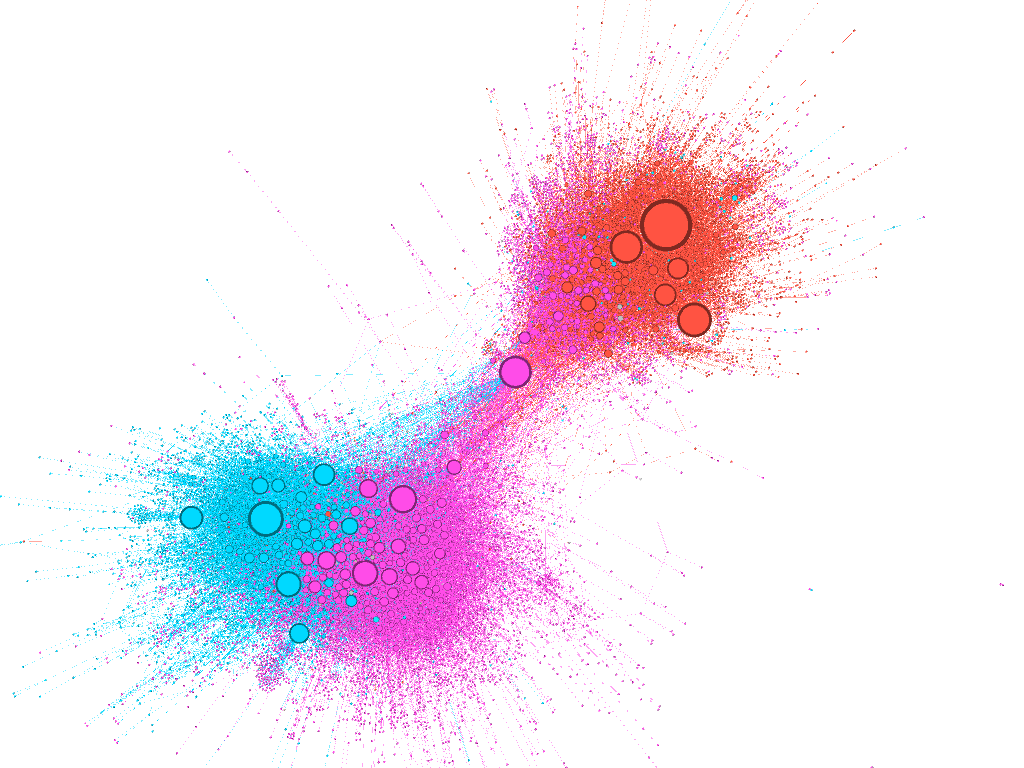}
  \caption{CAA/NRC}
  \label{caanrc_rt_graph}
\end{subfigure}
\begin{subfigure}{.5\columnwidth}
  \centering
 \includegraphics[width=0.99\linewidth]{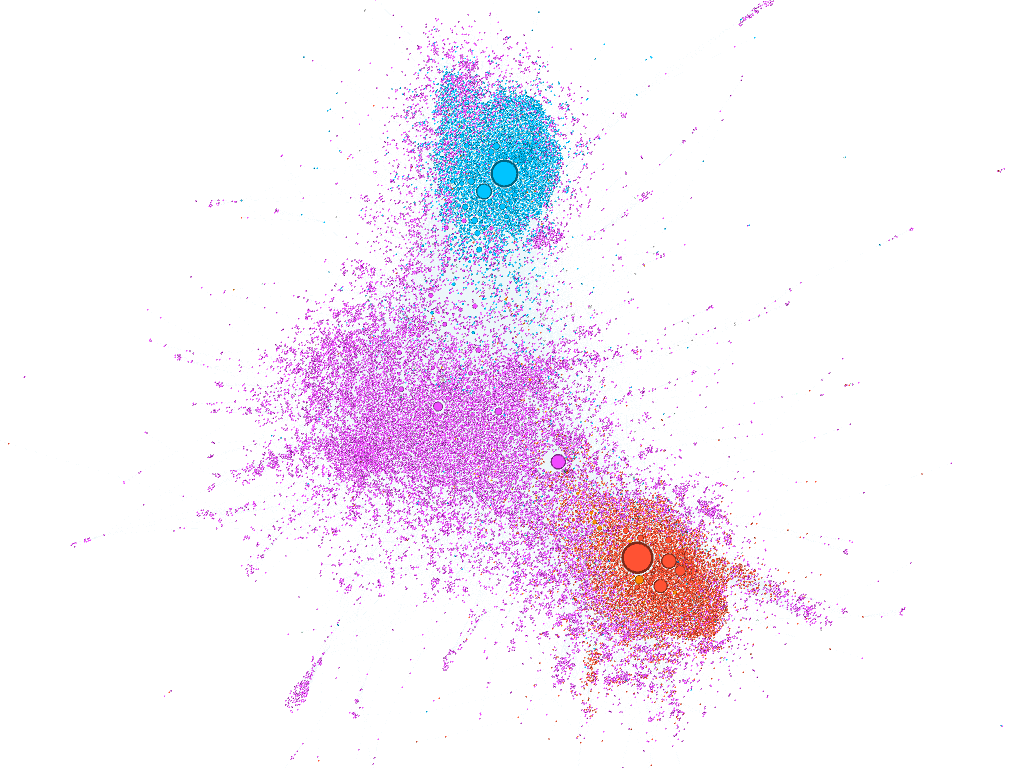}
  \caption{COVID-19}
  \label{covid_rt_graph}
\end{subfigure}%
\begin{subfigure}{.5\columnwidth}
  \centering
 \includegraphics[width=0.99\linewidth]{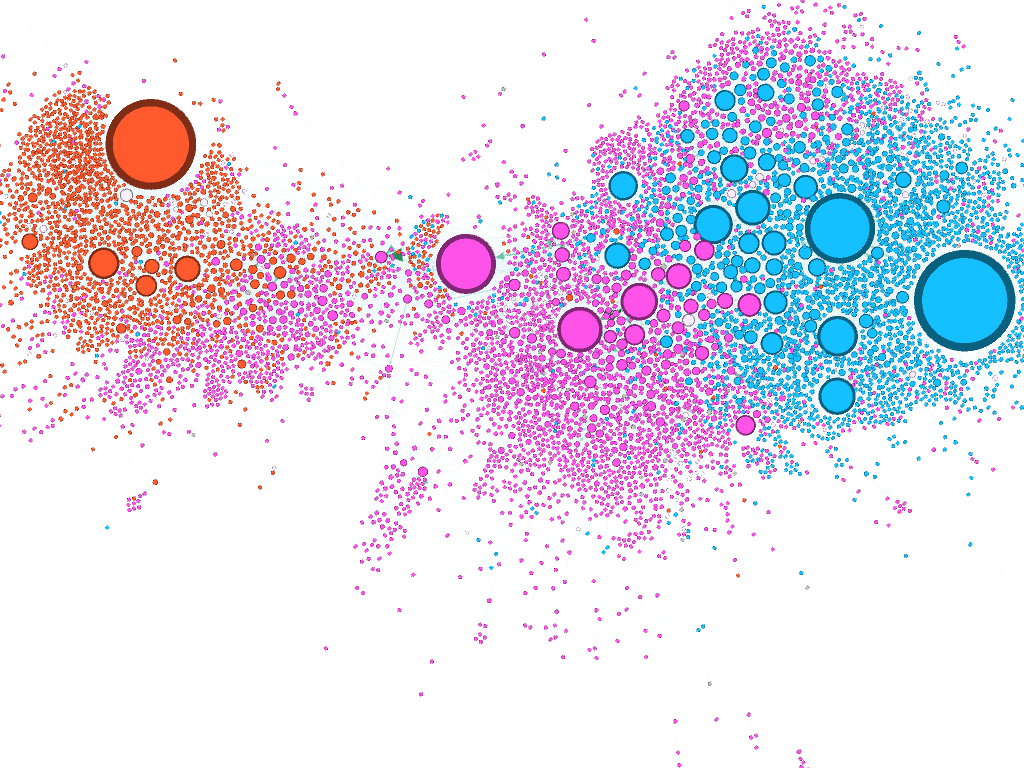}
  \caption{Farmers' Protests}
  \label{farmers_rt_graph}
\end{subfigure}

\caption{\textbf{Retweet Graph by Events}. BJP in orange, INC in blue, Influencers in pink. Partisan engagement of Influencers across all events in is evidenced by the network graphs.}
\label{event_rt_graph}
\end{figure}

The constructed retweet graph for all the events is shown in Figure \ref{event_rt_graph}.  Each node represents users who are either politicians or influencers and the edges represent retweets between them. The graph is color partitioned as INC in blue, BJP in orange and influencers in pink. The graph is visualised using the Force Atlas 2 layout in Gephi \cite{gephi2009}, where we set gravity rules for node attraction using their edge strength to further clarify the clusters. The size of the nodes indicates their degree centrality. Overall, similar to the clusters above, we observe that for all events with the exception of COVID-19, most influencers engage with INC and a small fraction of them engage with BJP. In the case of COVID-19, maximum engagement of influencers is with the incumbent party, BJP. 
Among the influencers, we observe that journalists, fan accounts and media houses occupy central positions in the network.  
\newline \indent
To compute the retweet polarity score, consider $\mathbf{X}^*$ and $\mathbf{Y}^*$ to be  the set of top $k$ (=100) high degree vertices, i.e. the top $k$ users with the most number of retweet edges, from partitions $\mathbf{X}$ and $\mathbf{Y}$ respectively. Let $l_u^X$ as defined in \citet{garimella2018quantifying} be the expected hitting time for a vertex $u \in \mathbf{V}$, i.e. the expected number of steps for a random walk starting with vertex $u$ and ending in a high degree vertex in $\mathbf{X}^*$. Therefore $p(u)^X$ is defined as the fraction of vertices $v \in \mathbf{V}$ for which the expected hitting time is more than that of vertex $u$, i.e. $l_u^X < l_v^X$. Therefore a score of $r(u)^X \approx 1$ indicates that the vertex $u$ is close to a high degree vertex in $\mathbf{X}^*$. Subsequently, $r(u)^Y$ is defined similarly, leading to retweet polarity score $r(u)$ being defined as:
\begin{equation}
    r(u) = r(u)^X - r(u)^Y \quad \in (-1,1)
\end{equation}
where a score close to 1 indicates that a user is more likely to be retweeted by another user with a pro-INC stance, while a score close to -1 indicates that the user is more likely to be retweeted by a user with a pro-BJP stance.

\subsubsection{Content Polarity}
We adopt the cultural axis generation procedure described by \citet{waller2020community}. We compute a partisan axis in the high dimensional user embedding space, by taking the difference of vector embeddings of the top $n$ (=10) INC and BJP politicians, sorted by retweet polarity scores. Briefly, let $\tilde{x}$ and $\tilde{y}$ be the high dimension embedding vectors representing INC and BJP politicians respectively. For any user $u$, with a high dimension embedding vector $\tilde{u}$, the content polarity score $c(u)$ is defined as the cosine similarity of the user with the partisan axis:
\begin{equation}
    c(u) = \frac{\tilde{u} \cdot (\tilde{x} - \tilde{y})}{\norm{\tilde{u}}\norm{\tilde{x} - \tilde{y}}} \quad \in (-1,1)
\end{equation}
where a score close to 1 indicates a user's stance on the particular issue is highly pro-INC, whereas a score close to -1 signifies that the user's stance on the issue is highly pro-BJP.

\subsubsection{Distribution of Polarity Scores}
\begin{figure*}[!htb]
\centering
\begin{subfigure}{.25\textwidth}
  \centering
  \includegraphics[width=0.97\linewidth]{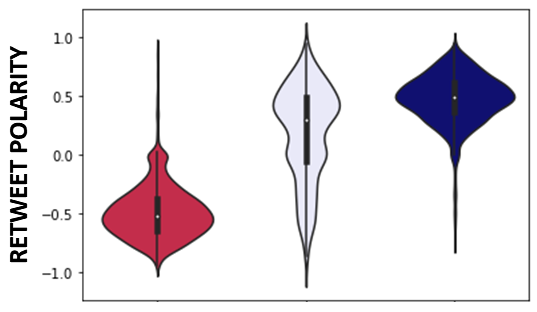}
  \caption{}
\end{subfigure}%
\begin{subfigure}{.25\textwidth}
  \centering
 \includegraphics[width=0.9\linewidth]{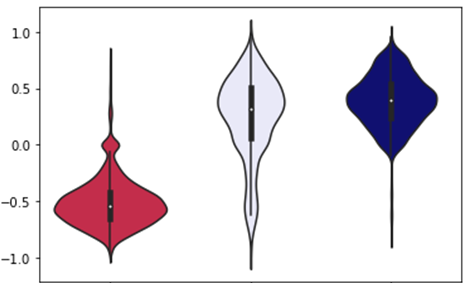}
  \caption{}
\end{subfigure}%
\begin{subfigure}{.25\textwidth}
  \centering
 \includegraphics[width=0.9\linewidth]{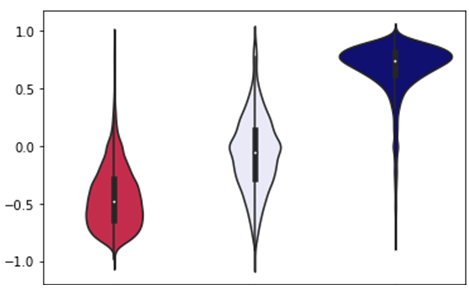}
  \caption{}
\end{subfigure}%
\begin{subfigure}{.25\textwidth}
  \centering
 \includegraphics[width=0.9\linewidth]{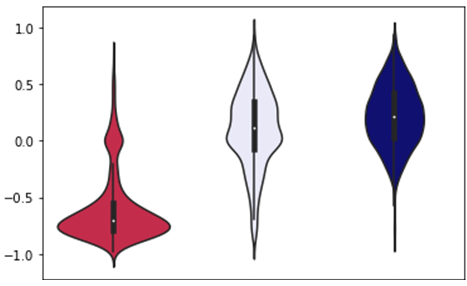}
  \caption{}
\end{subfigure}
\begin{subfigure}{.25\textwidth}
  \centering
  \includegraphics[width=0.97\linewidth]{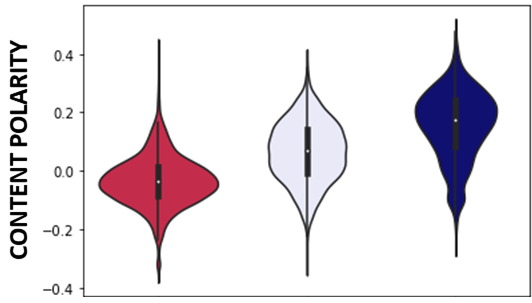}
  \caption{}
\end{subfigure}%
\begin{subfigure}{.25\textwidth}
  \centering
 \includegraphics[width=0.9\linewidth]{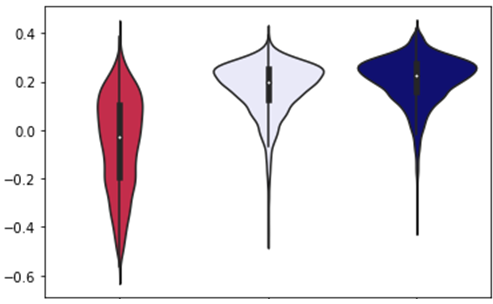}
  \caption{}
\end{subfigure}%
\begin{subfigure}{.25\textwidth}
  \centering
 \includegraphics[width=0.9\linewidth]{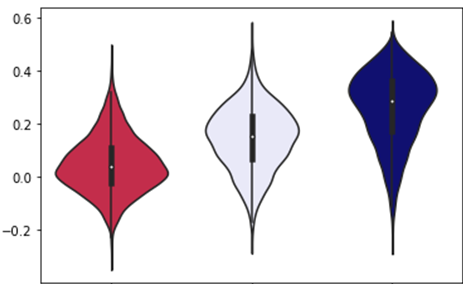}
  \caption{}
\end{subfigure}%
\begin{subfigure}{.25\textwidth}
  \centering
 \includegraphics[width=0.9\linewidth]{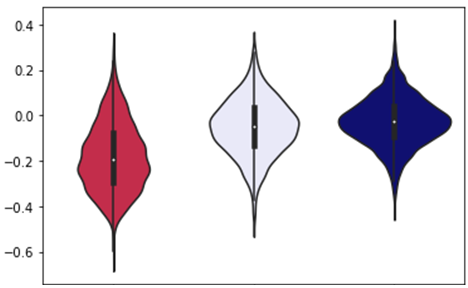}
  \caption{}
\end{subfigure}
\begin{subfigure}{.4\textwidth}
  \centering
 \includegraphics[width=0.9\linewidth]{figures/clusters_legend.jpg}
  \caption{}
  \label{legend}
\end{subfigure}

\caption{\textbf{Distribution of Polarity Scores by Event}. Subfigures (a) - (d) depict retweet polarity scores and (e) - (h) depict content polarity scores for Article 370, CAA/NRC, COVID-19 and Farmers' Protests respectively. The width of the violin plots represent the probability density of the data at different points, obtained from a kernel density estimator.}
\label{violin_plots}
\end{figure*}

Based on the clusters and retweet graphs from above, we calculate retweet and content polarity scores for all the influencers for a particular event. The distributions of the polarity scores in Figure \ref{violin_plots} shows that for COVID-19 the median scores for influencers is closer to the median of BJP politicians, while for Article 370, CAA/NRC and Farmers' Protest it is closer to the median of INC politicians. We also note that a significant number of influencers are consistently polarized towards a single party. This spurs our investigation on the incentives and characterizations of these hyper-partisan influencers.

\section{Rewards of Polarized Engagement}
In this section we use the computed polarity scores to investigate whether the partisan engagement of influencers is rewarded on Twitter, in terms of increased following and retweets. We assign an aggregate polarity score to an influencer by taking the median of the absolute polarity scores across all events.  
It should be noted that we use the absolute values of the polarity scores because we are interested in how the magnitude of the scores vary with attributes like number of followers etc. 

\subsection{Number of Followers}
To analyse whether increased polarization correlates with high following, we categorize the influencers by their number of followers. We categorize an influencer as \textit{Very Low} if the number of followers lies in the first quartile of the followers count distribution across all influencers, \textit{Low} if it lies in the second quartile, \textit{Medium} if it lies in the third quartile and High in the fourth quartile.

\begin{figure}[!htbp]
    \centering
    \includegraphics[width = 0.7\columnwidth]{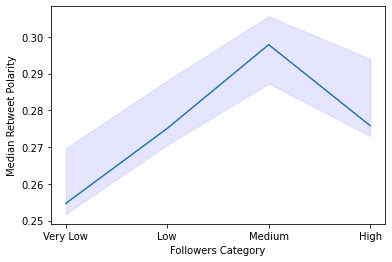}
    \caption{\textbf{Median Retweet Polarity Scores By Followers Category.} We observe that influencers with \textit{Medium} number of followers have the highest median retweet polarity score.}
    \label{rt_pol_fol}
\end{figure}

We then plot the median retweet polarity scores, along with the confidence intervals (derived from one-way ANOVA used below) for each of the categories in Figure \ref{rt_pol_fol}, where we see that the retweet polarity score increases with the number of followers. However, we see a marginal decrease in the polarity scores between \textit{Medium} influencers and \textit{High} influencers. A one-way ANOVA across scores of all categories results in a p-value $< 0.01$ and F statistic of 9.59. A post-hoc Honest Significant Difference (HSD) / Tukey's test reveals statistically signifcant differences between \textit{Medium} and \textit{Very Low} (p-value $< 0.01$), \textit{Medium} and \textit{Low} (p-value $< 0.1$) but no difference between \textit{Medium} and \textit{High}, indicating that the retweet polarity scores are significantly higher for \textit{Medium} to \textit{High} influencers as compared to influencers with lesser number of followers. 

\begin{figure}[!htbp]
    \centering
    \includegraphics[width = 0.7\columnwidth]{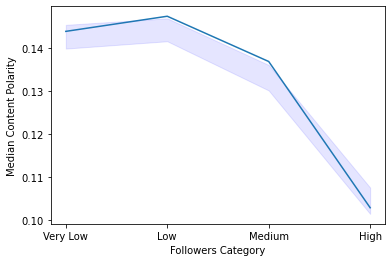}
    \caption{\textbf{Median Content Polarity Scores By Followers Category.} Influencers with \textit{Very Low} and \textit{Low} number of followers have the highest content polarity and those with \textit{High} number of followers have the lowest content polarity.}
    \label{content_pol_fol}
\end{figure}

We also plot the median of content polarity scores for each category in Figure \ref{content_pol_fol}, where we observe that \textit{Very Low} and \textit{Low} influencers have the highest content polarity. We then conduct a one-way ANOVA test for scores across all categories, which results in a p-value $< 0.01$ and F statistic of 157.41. Tukey's Test reveals that while \textit{Low} and \textit{Very Low}  are not statistically different (p-value $> 0.1$), \textit{Low} is significantly different from \textit{Medium} and \textit{High} (p-value $ < 0.01$).
\newline \indent
We observe that there is a \textit{twilight zone} for influencer polarization with respect to number of followers, wherein influencers with a certain range of followers are significantly more polarizing than other groups. In terms of retweet polarity, which quantifies how likely an influencer's tweet is to be retweeted by either side, influencers with \textit{Medium} followers have the highest polarity scores. Whereas in terms of content polarity, which signifies how extreme an influencer's stance is, those with \textit{Very Low} and \textit{Low} followers seem to have the highest scores. This suggests that influencers who are retweeted in a partisan manner generally have a moderate to high following and those who have an extreme stance on political issues are likely to have a low following. 
\newline \indent
Moreover, we find that from 20 of the influencers with the highest number of followers - mostly entertainers and sportspersons - only two were retweeted by the INC. This confirms suggestions in past work that the ruling party has a much stronger grip on the social media output of A-list celebrities \cite{lalani2019appeal}.


\subsection{Number of Retweets}
We fit a regression model with the following equation --
\textit{log\_retweetCount} $\sim$ \textit{polarityScore} $+$ \textit{log\_followersCount}, where \textit{polarityScore} is either \textit{retweetPolarity} or \textit{contentPolarity} and \textit{log\_followersCount} is used to control for \textit{log\_retweetCount}. The results are displayed in Table \ref{rt_count_reg}. In general, the median retweet rate is higher for influencers with higher polarity scores, potentially implying that polarized influencers are retweeted more in comparison to others.
To explore this further, we filter out influencers whose retweet polarity scores lie in the fourth quartile. We then calculate the median retweet rate for their polarizing tweets and median retweet rate for the rest of their tweets. We observe that 84\% of the influencers have a higher median retweet rate for tweets related to polarizing events as compared to their other tweets, potentially implying that polarized tweets are more likely to be rewarded with higher retweets as compared to other tweets.

\begin{table}[!htb]
\centering
\small
\begin{tabular}{l|c|c|c}
\multicolumn{1}{c|}{} & log\_retweetCount & $R^2$ & p-value                   \\ \hline
\rule{0pt}{2.6ex} retweetPolarity          & 1.93 (0.149) & 0.136     & \textless 0.001 \\
contentPolarity     & 0.2973 (0.491)        & 0.111               & \textless 0.001       
\end{tabular}
\caption{\textbf{Regression Analysis.} Log of Median Retweet Count (log\_retweetCount) is the dependent variable. The regression coefficient with the standard error, $R^2$ values and p-values are displayed. All results are statistically signifcant. }
\label{rt_count_reg}
\end{table}

\section{Influencer Polarization by Category}
We analyse how specific groups of influencers are consistently aligned towards a political party across all issues, and find that platform celebrities and fan accounts are usually aligned with the government (BJP) while journalists engage most with the opposition party INC.
\begin{figure*}[!htbp]
\centering
\begin{subfigure}{.5\textwidth}
  \centering
  \includegraphics[width=0.75\linewidth]{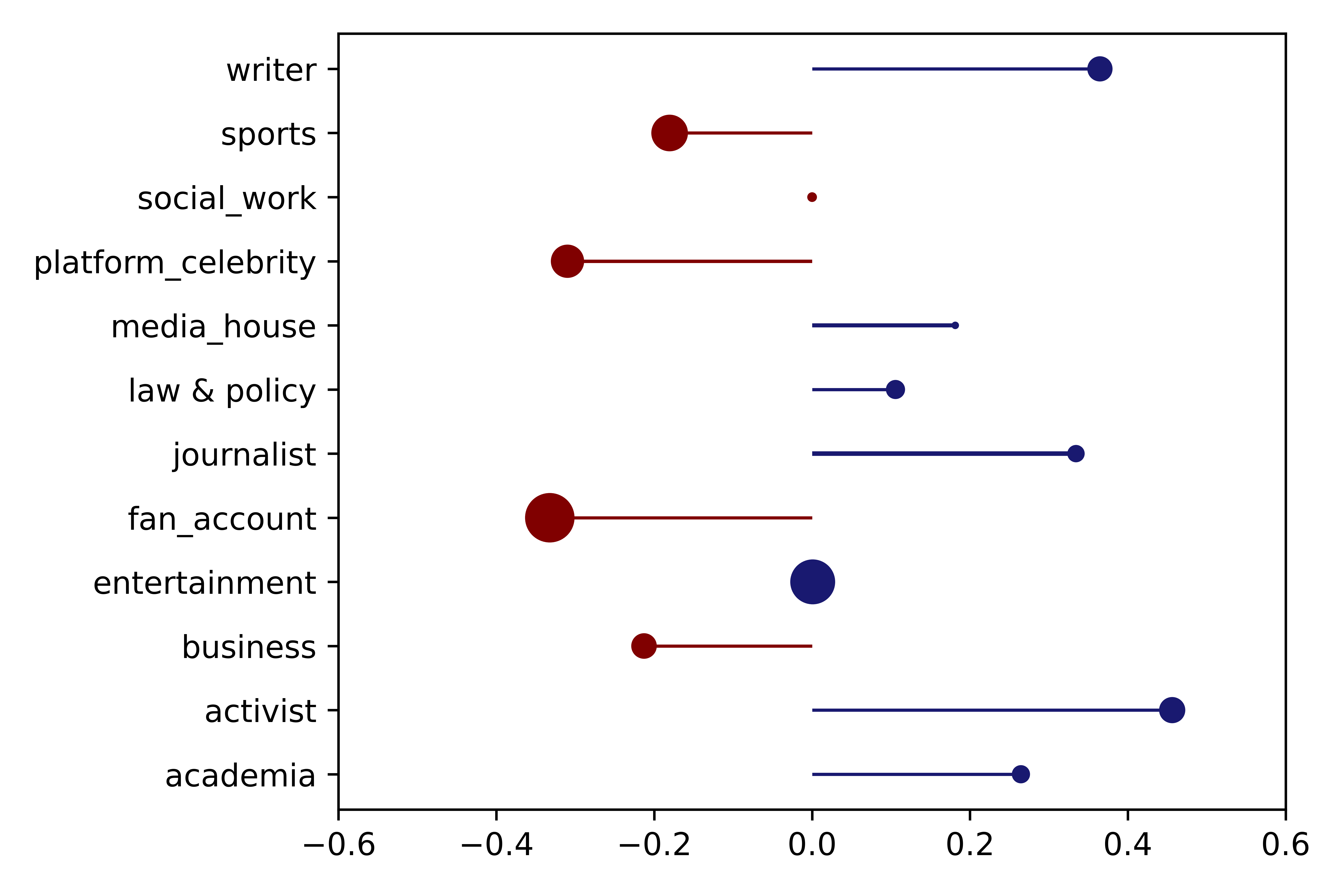}
  \caption{Article 370}
\end{subfigure}%
\begin{subfigure}{.5\textwidth}
  \centering
 \includegraphics[width=0.75\linewidth]{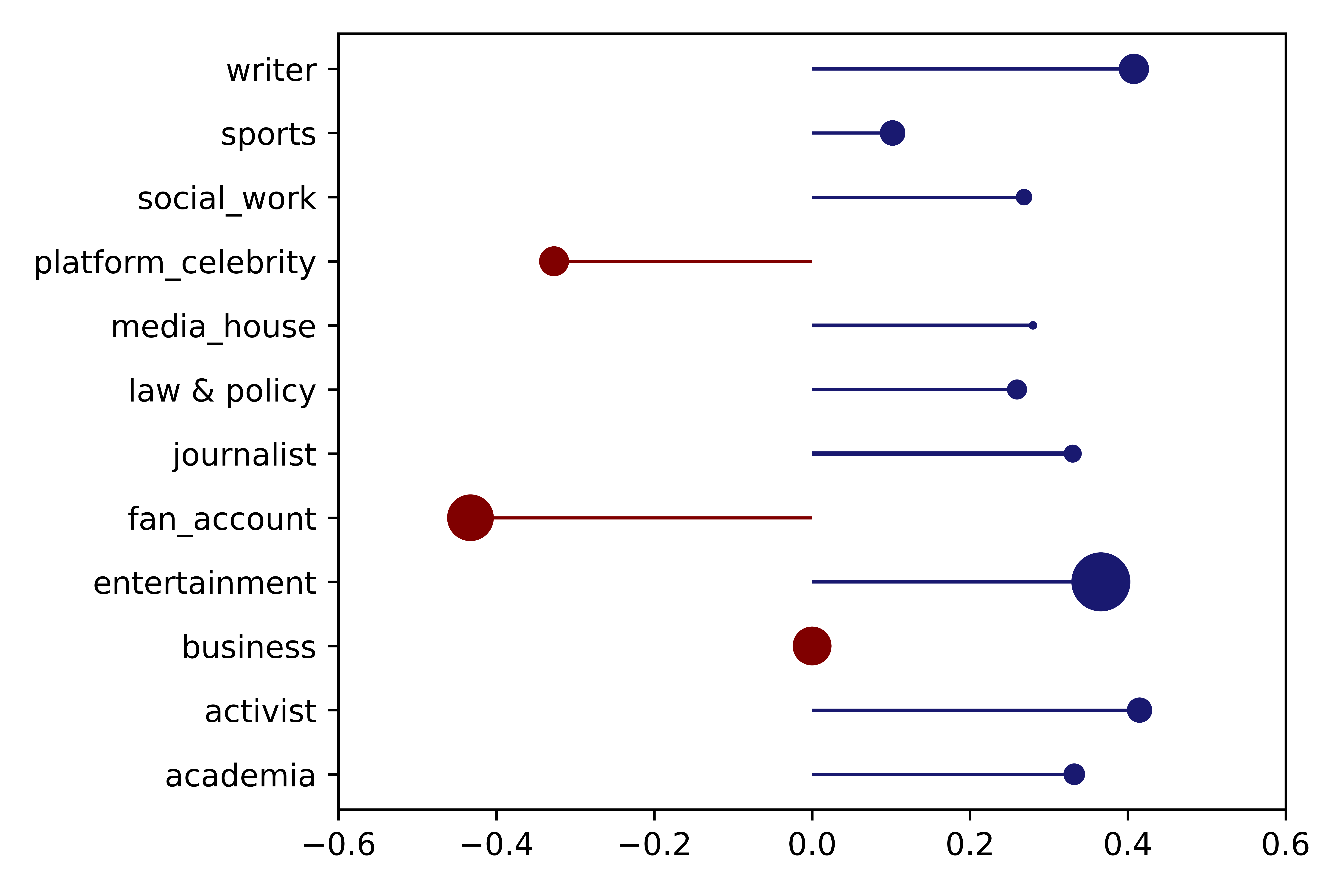}
  \caption{CAA/NRC}
\end{subfigure}
\begin{subfigure}{.5\textwidth}
  \centering
  \includegraphics[width=0.75\linewidth]{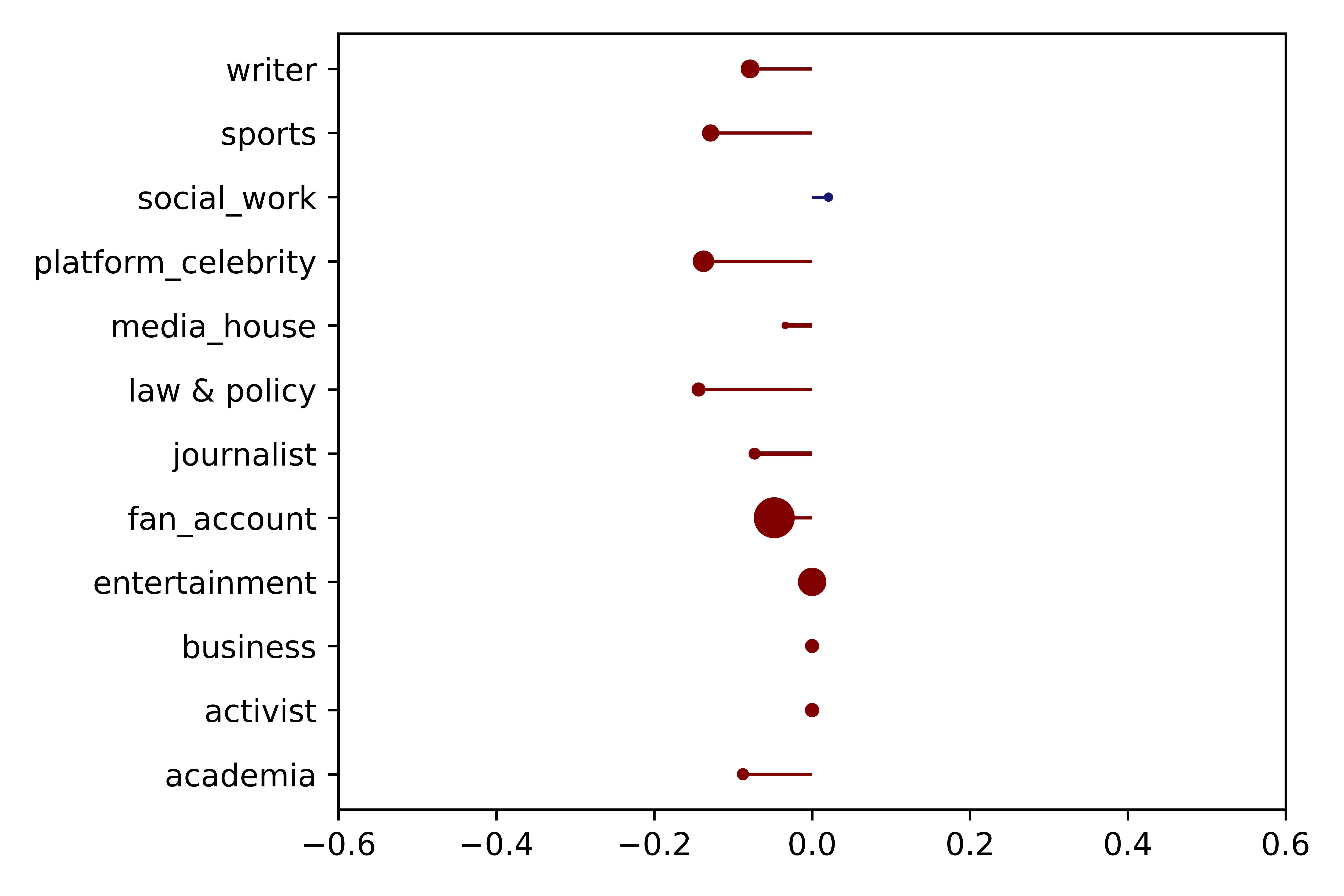}
  \caption{COVID-19}
\end{subfigure}%
\begin{subfigure}{.5\textwidth}
  \centering
 \includegraphics[width=0.75\linewidth]{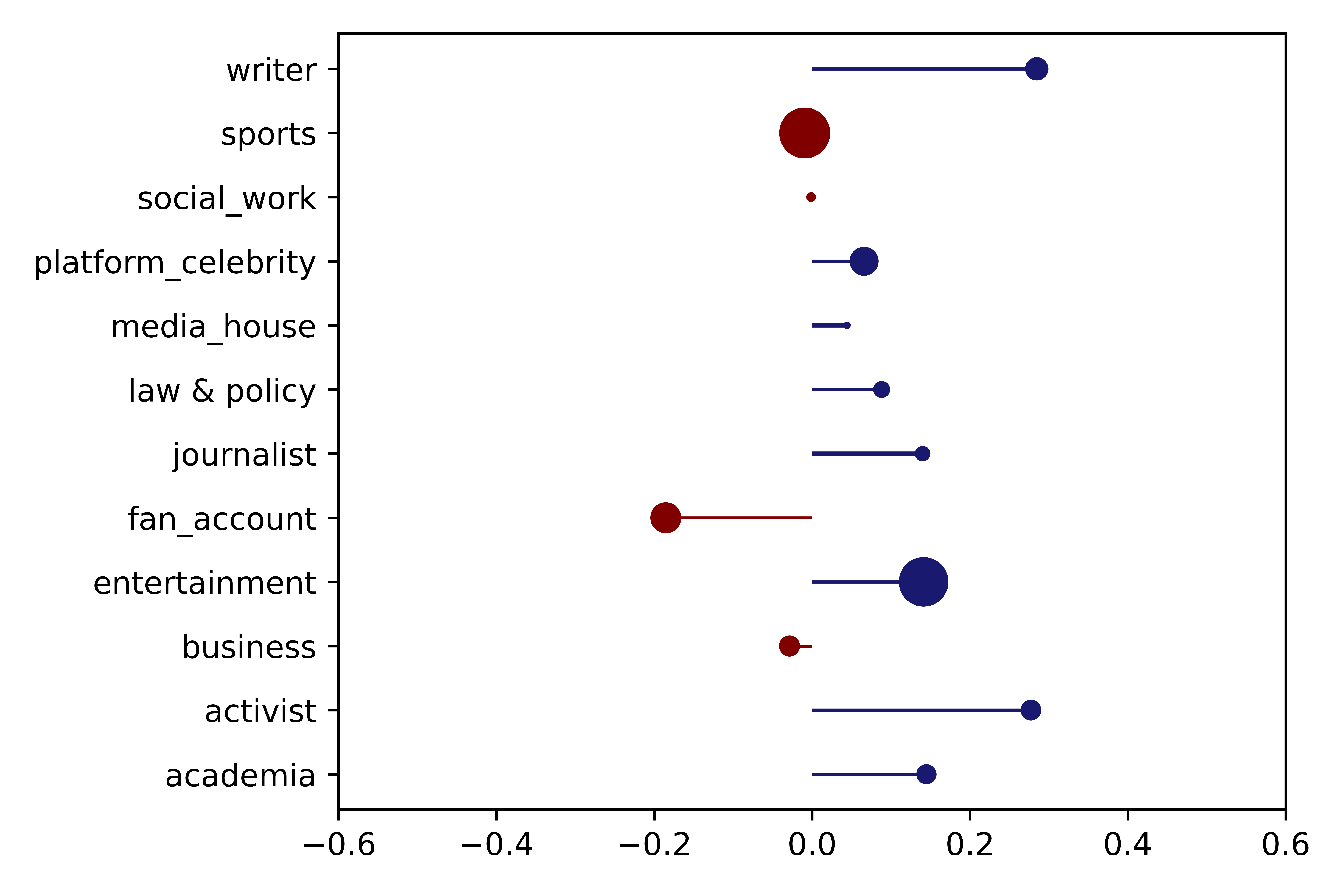}
  \caption{Farmers' Protests}
\end{subfigure}
\caption{\textbf{Median Retweet Polarity Scores by Influencer Category.} Length of stem indicates retweet polarity score and bubble size indicates median retweet count.}
\label{rt_polarity_cat}
\end{figure*}

\begin{figure*}[!htbp]
\centering
\begin{subfigure}{.5\textwidth}
  \centering
  \includegraphics[width=0.75\linewidth]{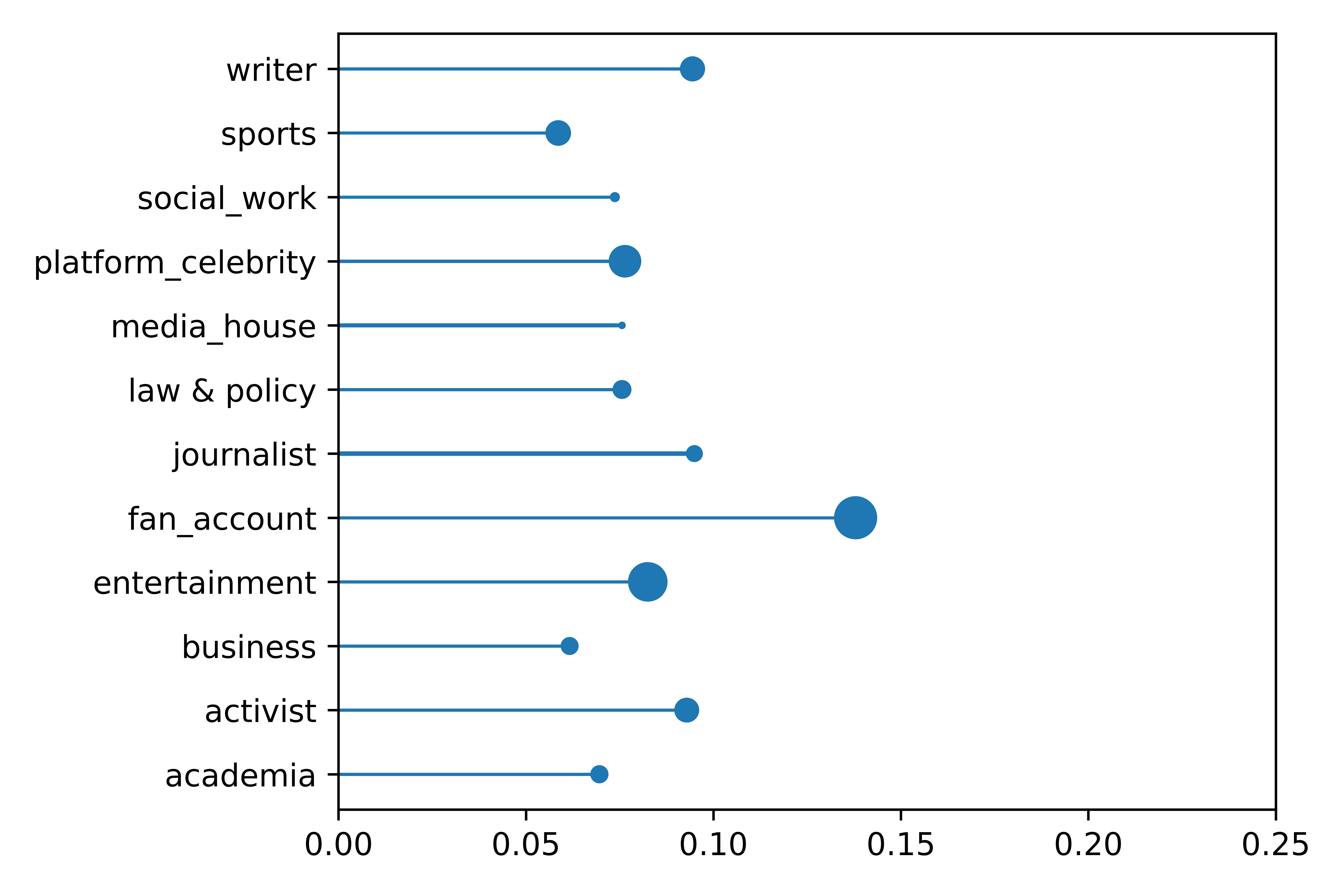}
  \caption{Article 370}
\end{subfigure}%
\begin{subfigure}{.5\textwidth}
  \centering
 \includegraphics[width=0.75\linewidth]{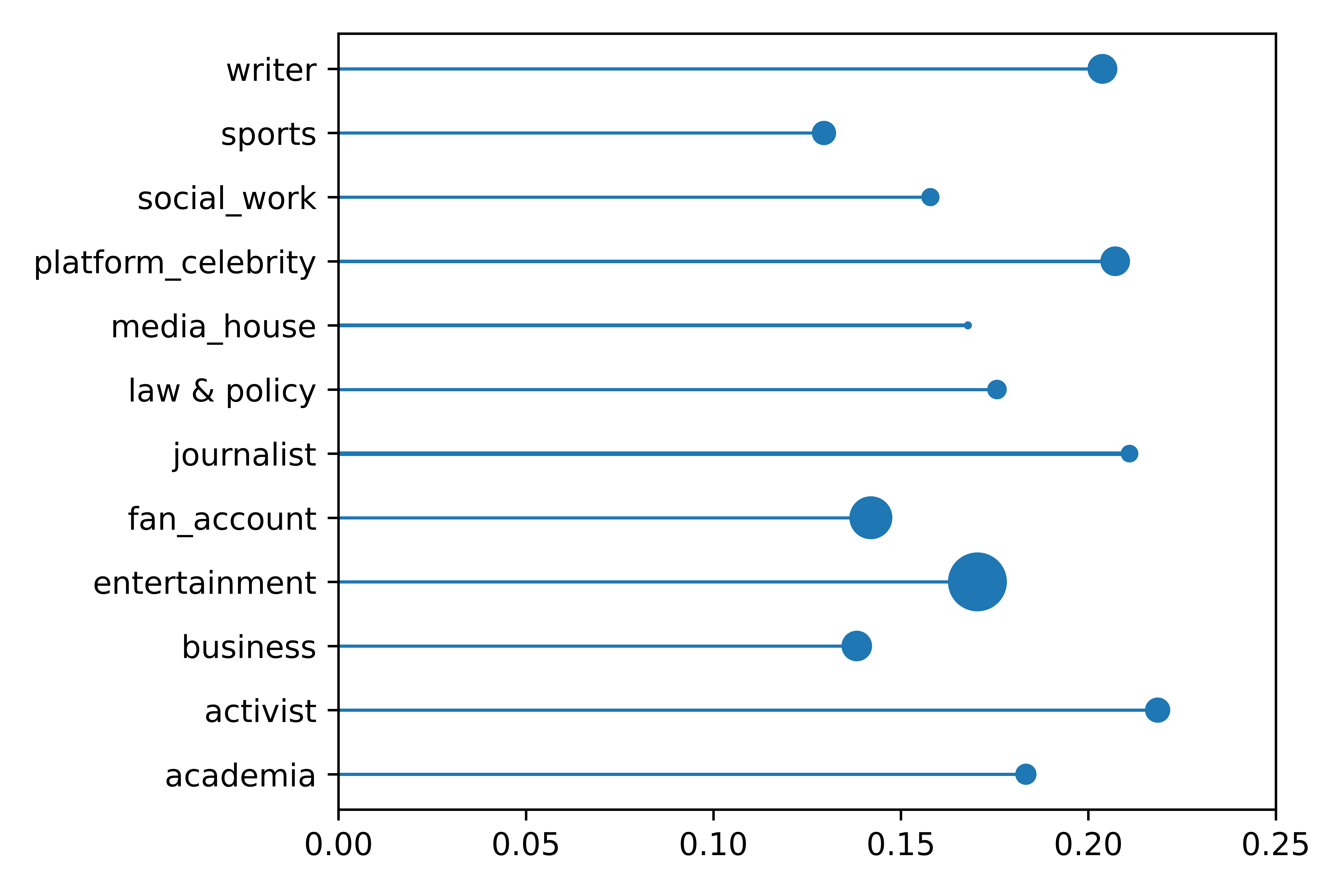}
  \caption{CAA/NRC}
\end{subfigure}
\begin{subfigure}{.5\textwidth}
  \centering
  \includegraphics[width=0.75\linewidth]{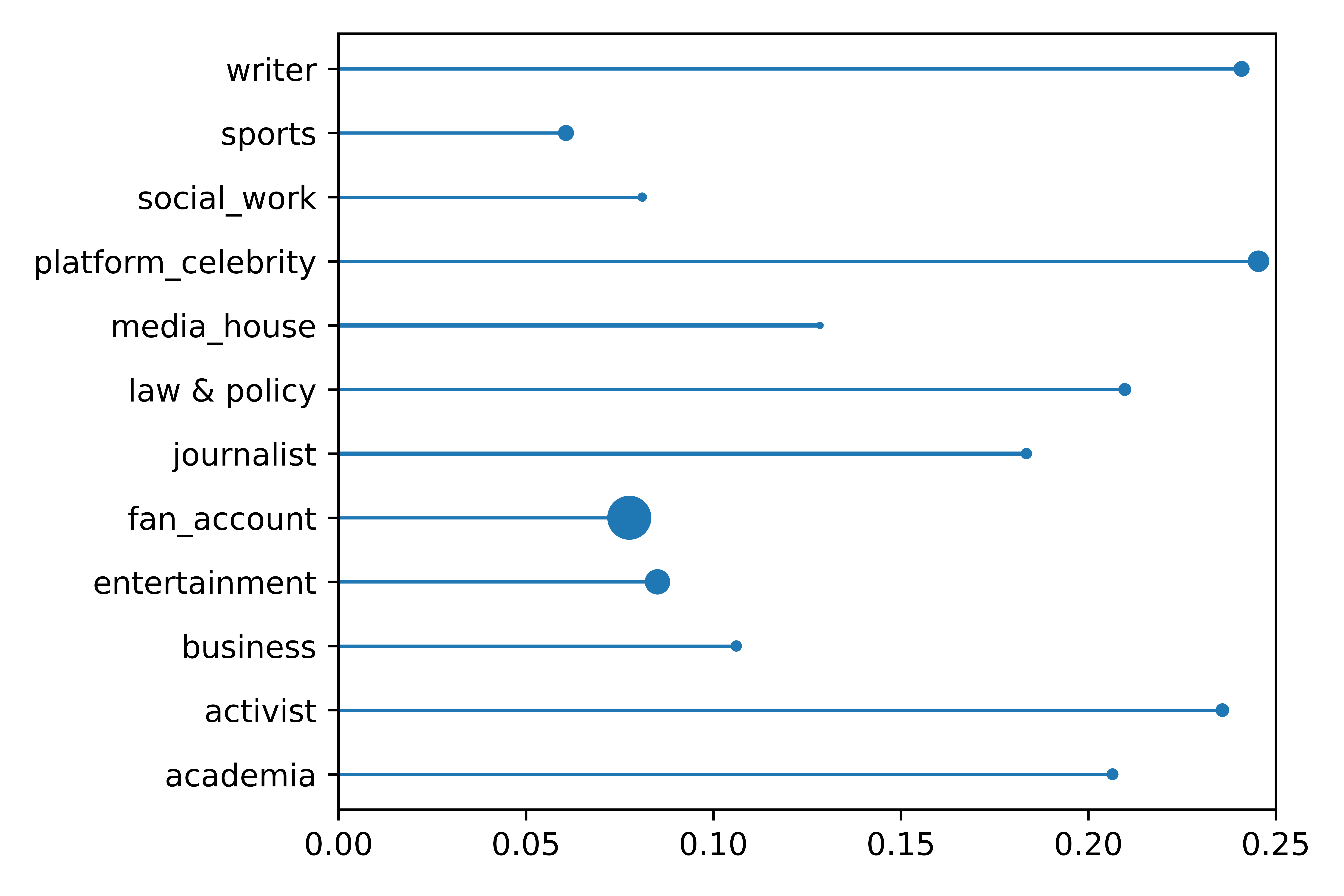}
  \caption{COVID-19}
\end{subfigure}%
\begin{subfigure}{.5\textwidth}
  \centering
 \includegraphics[width=0.75\linewidth]{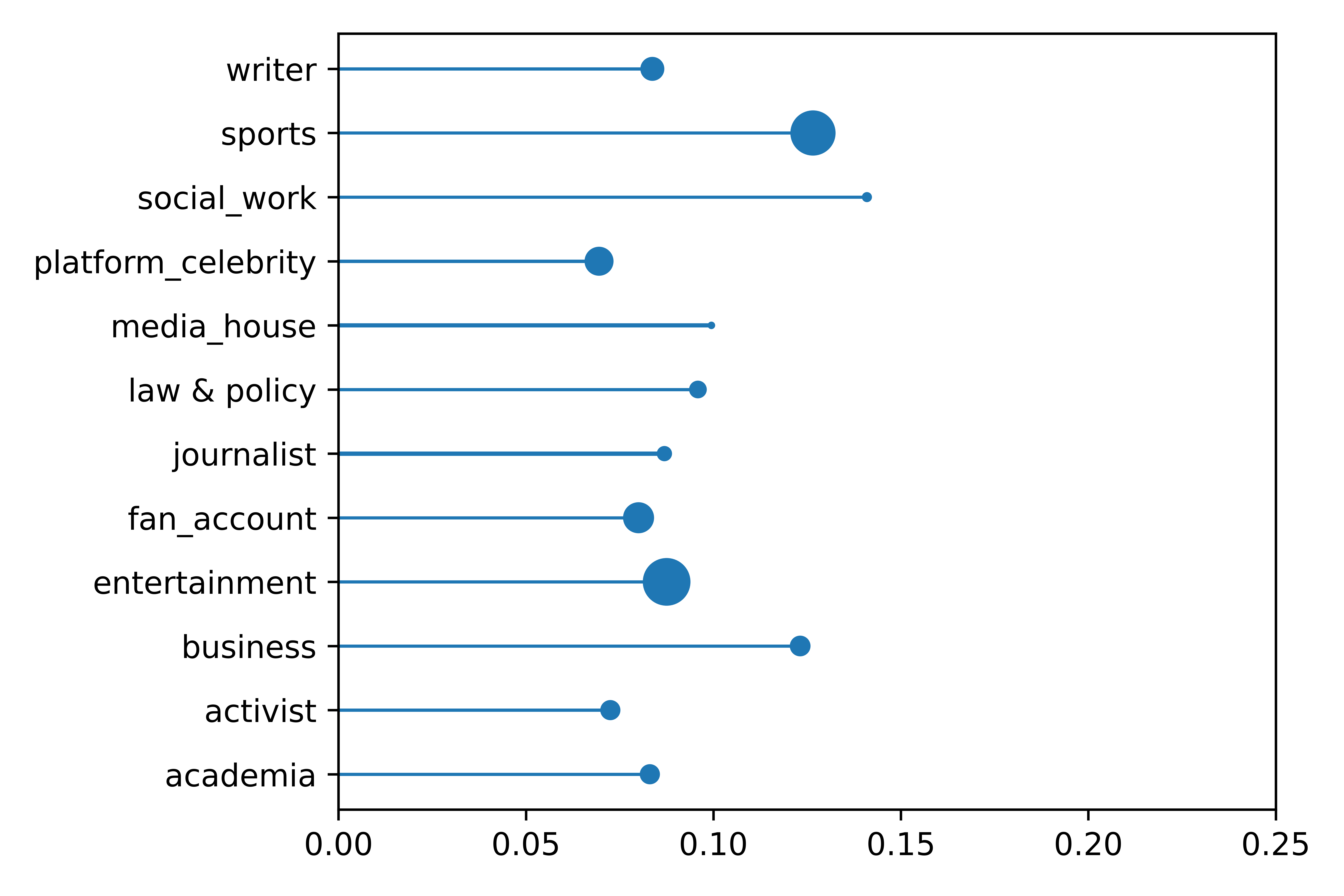}
  \caption{Farmers' Protests}
\end{subfigure}
\caption{\textbf{Median Content Polarity Scores by Influencer Category.} Length of stem indicates absolute content polarity score and bubble size indicates median retweet count.}
\label{content_polarity_cat}
\end{figure*}

\subsection{Retweet Polarity}
For each of the four events, we show in Figure \ref{rt_polarity_cat} how the retweet polarity scores vary with the category of the influencers. A close examination of the top 20 polarized accounts towards either side in each category reveals that, barring COVID-19, the INC side overwhelmingly endorses journalists on the three political issues. 
\newline \indent
For instance, in the activity after the abrogation of 370, 18 of the top 20 accounts with polarity towards the party are journalists. The BJP’s most proximate accounts have varied profiles in comparison, including platform celebrities, businesses, fan accounts, sports, and journalists. Unlike other issues in which voices on the ground of journalists, academics etc. are key to the political discourse, on the 370 issue, there is no representation of Kashmiris on-ground among top accounts on either side. 
This absence can partly be attributed to the shutdown of internet services and jailing of Kashmiri politicians, but nonetheless points to the lack of representative voices in the discourse. 
\newline \indent
Except journalists and activists, other categories of influencers are less likely to take stances critical of the government. We speculate that certain professions are more hesitant to be critical and are thus disproportionately retweeted by the BJP side. Here, de-politicized fields such as sports, businesses and entertainment have also shown tilts towards the ruling government. The entertainment category polarizes towards INC only during the CAA protests, which is in step with the on-ground developments in the movement when film industry figures spoke out against the Modi government's for the first time in significant numbers.

The lower ratio of journalists in the BJP’s top polarity scorers is an interesting trend that has connections to the nature of Modi’s political campaign. As \citet{chakravartty2015mr} note, in Modi's populist rhetoric, liberal media has remained as a prominent antagonistic signifier, ever since the press covered his role in the 2002 Gujarat riots where Muslims died in disproportionate numbers. This aversion, as our study shows, permeates to the ruling party's social media strategy after coming to power. Of the top media-related accounts the party centres, many are linked to partisan media houses, especially online portals that have gained prominence as vocal defendants of Hindutva ideology. Our finding is relevant with the \citet{recuero2020hyperpartisanship} study that observed the distancing of hyper-partisan Twitter clusters in Brazil from mainstream media. With INC, we observe that the top accounts are not as much supportive of the party as they are critical of the ruling government. 

The dissimilar polarity scores on the COVID-19 crisis also indicate a unique set of responses it set off among Twitter influencers. We attribute the shift in polarity scores towards the BJP, to the change in the type of the crisis, as pandemic foremost raised concerns about public health and preparedness, which is a bipartisan issue.
Here it is likely, that rather than being critical actors that invite accusations of being dispiriting, influencers were engaged in raising morale as part of government's campaigning and disseminating frontline information and resources. We also see journalists and media houses disproportionately amplified by the BJP, indicating a trend of relative cooperation with the government in the face of a global health emergency. 

\subsection{Content Polarity}
A category-wide content polarity analysis in Figure \ref{content_polarity_cat} for the four events produces distinct results. Of the four issues, we see that Article 370 draws some of the most uniform scores between categories, that are lower on content polarity. Kashmir as a territory is considered central to India’s narratives of national security and territorial claims. Historically, conversations about the region are sensitive and regulated, attracting legal prosecution \cite{bose2009kashmir}. This context seems relevant in understanding these scores. 
Within these relatively low scores, the top accounts are journalists and academics  critical of the move. That the high score on the government's side is driven by fan accounts and platform celebrities calls to question the role of hyper-partisan voices, who may not have on the ground knowledge of the topic.

CAA/NRC protests score high on content polarity in comparison. Of the top 20 polarizing accounts, an overwhelming proportion has to do with journalism, including reporters and media houses, largely those critical of the government. We see the presence of Indian public service broadcasters (e.g  @DDNational, @DD\_Bharati etc.), which in principle, are meant remain autonomous and neutral. Content from these accounts have relayed the official line and are thus engaged significantly by  BJP politicians. 
COVID-19 has had the most mixed reaction between the categories in terms of the content, possibly because on the whole it is a less polarizing event. The relatively polarized accounts here are often explicitly aligned with one of another party. 

On the farmers' protests, the top 20 accounts are divided between supporters and opponents of the government with a cross-category representation. Here too, government-aligned accounts, in the form of public broadcasters and industrial consortia are prominent. Somewhat exceptionally, businesses, who protesters see as the real beneficiaries of the policy, have a high polarity score on this issue. Our analysis of tweets from the highest-scoring accounts in this category shows how business houses, while avoiding explicit political keywords, have used a sanitised, corporate vocabulary to convey the benefits of the farm bills. Sportspersons have also only had higher polarity scores on this topic, most likely due to a popular singer, Rihanna's tweet on the issue, which triggered several influencers to speak out in favour of the government \cite{mishra2021rihanna}. Note that we use the absolute values of content polarity scores for this categorical analysis, to yield interpretable results in the experiments.

\section{Discussion}

In this work, we present a non-western case of partisan online engagement in a polarized political environment. We propose a novel workflow that uses event-specific tweets, labelled using a word embeddings based bag of words model, to compute the stance of Indian Twitter influencers on politically polarizing issues. We find that the influencers engage on these issues in a partisan manner, favouring either the ruling party, BJP or the opposition party INC. We also propose retweet and content based polarity metrics to quantify this partisan engagement. We further use the polarity scores to characterize polarized influencers, and find that partisan influencers are retweeted and followed more than other users, underlining the fact that influencers are now an essential piece of this communication ecology. This is similar to the conclusions in \citet{jiang2021social}, where they find links between online influence and partisanship in the United States. 

Moreover, we show that the differences in clustering around topics are themselves insightful in understanding the larger patterns in the discourse. Focusing our examination around key events, helps understand the periods of elevated activity, i.e. those in which political parties may resort to engaging the various outreach resources at their behest, including influencers. In this work, we find that that the opposition gains tailwind from the work of journalists and commentators, who usually tend to dissent with the government, while fan accounts and platform celebrities are prominent in their polarisation towards the ruling party. This warrants further enquiries into the mobilisation of new types of influencers, to convey the appearance of political legitimacy. 
\newline \indent
While our findings are immediately of relevance in understanding the political economy of social media in India, they are also relevant to approaching the political engagement of influencers in other contexts. As discussed in \citet{bodrunova2019beyond}, contrary to most studies emerging from USA, our study also finds that influencers are not always polarized based on their political leaning. For instance, while the events of Article 370, CAA and Farmers' Protests, which were policy based issues, saw party-based polarization, there was a significant confluence of influencers on the side of the government during the COVID-19 crisis. This suggests that issue-based polarization might play a larger role in multi-ethnic democracies like India, than previously thought.
\newline \indent 
An existing limitation of our study is that we only consider English-language tweets. In a multilingual setting like India, tweets in regional languages are emerging as an indispensable source for understanding linguistically altered narratives \cite{srivastava2020understanding}. However, our methodology can be easily extended to the multilingual setting, by using a language specific bag of words model to label the tweets and then using the multilingual Universal Sentence Encoder trained on over 50 languages \cite{reimers2020making} to compute the stance of users.
\newline \indent 
Our study lays the ground for future work. There is an emerging need to study the various nuanced categories of influencers, their relationship with the government and their role in exacerbating political polarization online. While we studied the intersection of influencer and politician behavior during crises, it is equally important to extend such work into its typical daily patterns.  

\bibliography{ref.bib}

\end{document}